\documentclass[aps,notitlepage,twocolumn,nofootinbib,pra,10pt,floatfix,usenames,dvipsnames]{revtex4-2}
\usepackage{amsmath,amsfonts,amssymb,amsthm,bbm,graphicx,enumerate,times,mathtools,physics,marvosym,wasysym,soul,stmaryrd,mathrsfs}
\usepackage[usenames,dvipsnames]{xcolor}
\usepackage{float}
\usepackage{hyperref}
\usepackage{soul}

\hypersetup{colorlinks,linkcolor=teal,citecolor=teal,urlcolor=teal}
\usepackage[ruled,longend]{algorithm2e}
\usepackage{etoolbox}
\usepackage[noabbrev,capitalise]{cleveref}
\usepackage{comment}
\usepackage{dsfont}
\usepackage{stmaryrd}

\definecolor{alex}{rgb}{.2,.7,.1}

\newcommand{\ii}{\ensuremath\mathrm{i}}
\newcommand{\CC}[1]{\textcolor{blue}{(CC) #1}}

\theoremstyle{definition}

\begin{document}

\title{Universal fault-tolerant logic with heterogeneous holographic codes}
\author{M.~Steinberg$^{1,2,6}$, J.~Fan$^{1,2}$, J. Eisert$^{4}$, S.~Feld$^{1,2}$, A.~Jahn$^{3}$, C. Cao$^{5}$}
\affiliation{$^{1}$QuTech, Delft University of Technology, 2628 CJ Delft, The Netherlands}
\affiliation{$^{2}$Quantum and Computer Engineering Department, Delft University of Technology, 2628 CD Delft, The Netherlands}
\affiliation{$^{3}$Department of Physics, Freie Universit\"at Berlin, 14195 Berlin, Germany}
\affiliation{$^{4}$Dahlem Center for Complex Quantum Systems, Freie Universit\"{a}t Berlin, 14195 Berlin, Germany}
\affiliation{$^{5}$Department of Physics, Virginia Tech, Blacksburg, VA 24061, USA}
\affiliation{$^{6}$Global Technology Applied Research, JPMorganChase, New York, NY 10017 USA}

\begin{abstract}
The study of holographic bulk-boundary dualities has led to the construction of novel quantum 
error correcting
codes. 
Although
these codes have shed new light on conceptual 
aspects of these dualities, they 
have widely been believed to lack a crucial feature of practical quantum error correction: The ability to support universal fault-tolerant 
quantum
logic. In this work, we introduce a new class of holographic codes that realize this feature. These heterogeneous holographic codes are constructed by combining two seed codes in a tensor network on an alternating hyperbolic tiling. We show how this construction generalizes previous strategies for fault tolerance in tree-type concatenated codes, allowing one to implement non-Clifford gates fault-tolerantly on the holographic boundary. We also 
demonstrate that these codes allow for high erasure thresholds under a suitable heterogeneous combination of specific seed codes. Compared to previous concatenated codes, 
heterogeneous holographic codes achieve large overhead savings in physical qubits, 
e.g., a $21.8\%$ reduction for a two-layer Steane/quantum Reed-Muller combination. Unlike standard concatenated codes, we 
establish that 
the new codes can encode more than a single logical qubit per code block by applying ``black hole'' deformations with tunable rate and distance, 
while possessing fully addressable, universal fault-tolerant gate sets. Therefore, our work strengthens the case for the utility of holographic quantum codes 
for practical quantum computing.
\end{abstract}

\maketitle



\section{Introduction}\label{section:intro}

\emph{Quantum error correction} (QEC) remains thus far the only known strategy for building a large-scale, fault-tolerant quantum computer \cite{campbell2017roads,nielsen_chuang} in the presence of unavoidable noise. As a result, current research centers on practically all parts of the QEC code development cycle, including novel code constructions and families, estimations of error resilience for encoded memories, methods for implementing universal logic, \emph{fault-tolerant} (FT) protocols for syndrome extraction, architectural designs, and 
efficient classical decoding strategies that 
amount to find suitable correction strategies given data
from syndrome measurements \cite{lidar_qec_book,gottesman2022opportunities,terhal2015quantum}. Among these considerations, the assessment of universal fault-tolerant logical gate sets in light of the famous \emph{Eastin-Knill theorem}, a known restriction regarding the inadmissibility of universal transversal gates in QEC codes, is of paramount interest \cite{eastin_knill_theorem}. 

In principle, many methods have been studied for 
in effect 
bypassing the Eastin-Knill theorem. Examples of these techniques include \emph{magic-state distillation} \cite{bravyi2005universal,bravyi2012magic} 
that add suitable resource states that are being distilled to high quality and then used to implement universal quantum operations. Then,
\emph{pieceable fault tolerance} \cite{yoder_piece} has been considered, \emph{code switching} \cite{paetznick_code_switching,costuniversality,anderson2014fault_conversion}, constructions of \emph{heterogeneous concatenated codes} \cite{hetero_tree1,hetero_tree2,hetero_tree3}; \emph{gauge fixing} \cite{chamberland2017complementarygauge,bombin2015gauge,vuillot2019code}, and constructions of non-stabilizer codes with \emph{exotic} transversal logical gate sets \cite{kubischta2023family,kubischta2024permutation}. While all of these strategies describe how universal FT quantum computation at the logical level can be attained, each has its own respective tradeoffs with respect to additional gate depth, locality, and -- often excessively demanding -- spacetime overhead \cite{koenig_ftlogic,yoshida_ft_logic,baspin2022connectivity,delfosse2021bounds,baspin2023lower,dai2024locality}. All this 
signifies  
that the extensive study of FT logical gate sets is indispensable for the future viability of quantum computing.

As mentioned above, protocols which circumvent the Eastin-Knill theorem have been studied in the context of concatenated quantum codes \cite{hetero_tree1,hetero_tree2,hetero_tree3}. Here, a universal FT logical gate set has been constructed by concatenating the Steane code \cite{PhysRevLett.77.793}, whose transversal logical gates are within the Clifford group, together with 
the \emph{quantum Reed-Muller} (QRM) code \cite{Steane:1996ci}, which is known to possess transversal logical $T$ and $CCZ$ gates \cite{paetznick_code_switching,anderson2014fault_conversion,koutsioumpas2022smallest}. In spite of the reported benefits of achieving universal FT logic \cite{hetero_tree1}, it has been found in
Ref.\ \cite{hetero_tree2} that the FT threshold of the resultant concatenated code is quite low. Consequently, orders of magnitude more physical qubits would be required in order to approach the same logical error rate for performing a logical $H$ or $T$ gate as for the surface code \cite{hetero_tree3}.

Initially developed with a very
motivation in mind, the first \emph{holographic quantum error correction} codes
\cite{happy_code_paper,jahn2021topical}
have been devised in order to 
elucidate important aspects of the celebrated \emph{AdS/CFT correspondence}
in a precise and explicit fashion, a mathematical duality posited between a $d$-dimensional \emph{conformal field theory} (CFT) and a $(d+1)$-dimensional theory of quantum gravity \cite{maldacena1999large,witten1998anti,nuastase2015introduction,gubser1998gauge,ammon2015gauge}. 
Discrete models of holographic 
commonly rely on the usage of \emph{tensor networks} to capture quantum information aspects of 
AdS/CFT, which include the Ryu-Takayanagi formula for \emph{holographic entanglement entropy} \cite{ryu2006holographic,almheiri2015bulk,Jahn}, bulk recovery and \emph{entanglement wedge reconstruction} \cite{hubeny2007covariant,hamilton2006holographic,freivogel2016precursors,Bao:2016skw,dong2016reconstruction}, \emph{renormalization group flows} \cite{furuya2022real,chen2020entanglement}, and \emph{subsystem complementary recovery} \cite{harlow2017ryu,pollack2022understanding,cao2021approximateBScode}. 
Despite their success as toy models of the gauge/gravity duality \cite{jahn2021topical}, the motivation of bringing ideas of quantum error correction to those of holography has to date been mostly of a conceptual
nature. The feasibility of holographic tensor-network codes for FT quantum computation has been met with general apprehension,
in fact, it has not even been anticipated to be true or seen as a desirable
feature. One key reason for this unease lies in the belief that transversal logical gates in holographic codes must be restricted to the Clifford group, ergo limiting their appeal for magic-state distillation and FT quantum computing \cite{williamson_holographic_ft_logical_gates}. Even more generally, the lack of global symmetries \cite{harlow2021symmetries,constraintssymmetriesholography} and the non-triviality of area operators \cite{cao2024non} in quantum gravity further constrain the prospects for FT logical gates. This collection of insights seems to suggest that 
emergent gravity and fault tolerance are in tension.
\begin{figure*}
\centering
\includegraphics[width=0.95\textwidth]{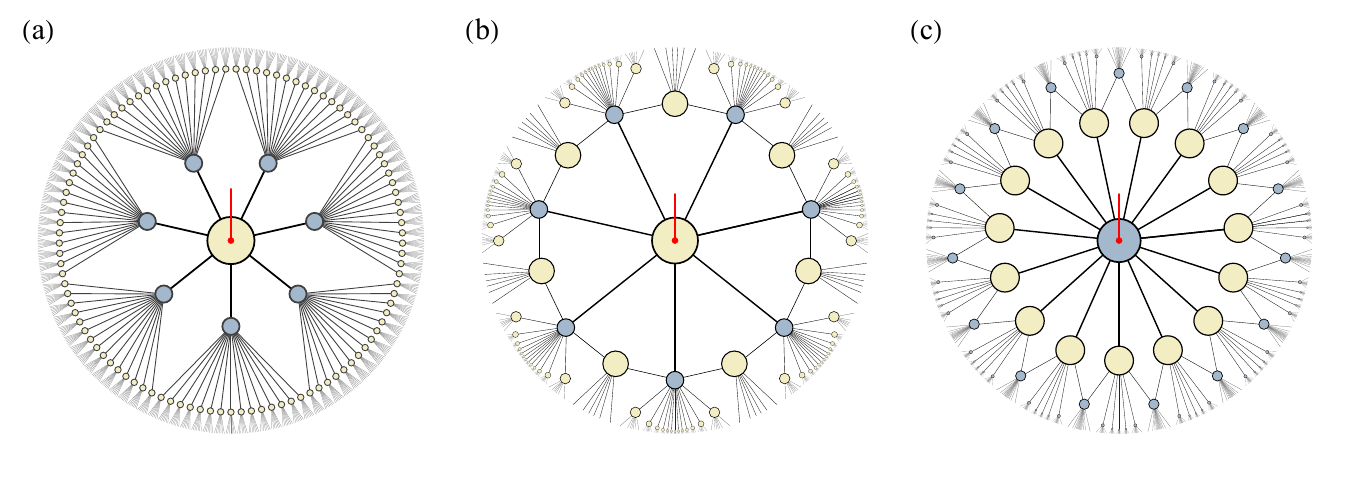}
\caption{Tensor networks defining heterogeneous holographic codes. (a) A Steane/QRM concatenated code as a tree tensor network for one logical qubit (red tensor leg), corresponding to a $\{16,\infty,8\}$ hyperbolic tiling. QRM code tensors are shown in dark blue and Steane code tensors in light yellow. The physical qubits of the code are identified with the black open legs on the tensor network boundary. 
(b) The alternating Steane/QRM code defined on a $\{8,2,16\}$ hyperbolic tiling. 
(c) Starting from a central QRM instead of Steane tensor leads to a QRM/Steane code on the equivalent $\{16,2,8\}$ hyperbolic tiling.
}
\label{fig:QRM.Steane.heterogeneous}
\end{figure*}
It has been shown in Ref.\ \cite{williamson_holographic_ft_logical_gates} that codes which even approximately adhere to the restriction of \emph{subsystem complementary recovery} can only possess transversal logical gates inside the Clifford group. The proof of this fact hinged on the assumption that the recoverable bulk algebra on a subregion always corresponds to a tensor product of qubits, and that subsystem complementary recovery can be upheld in all but a few cases. This assumption necessitates the construction of a tensor-network code \cite{cao_qlego,farrelly2021tensor} constructed uniformly from either tensors corresponding to \emph{absolutely maximally-entangled} (AME) quantum many-body states \cite{happy_code_paper,ame_states1,enriquez2016maximally}, or quantum states that are closely related, such as \emph{planar maximally-entangled} (PME) states \cite{pme_states}; these states are known to exhibit maximal entanglement entropy over various restrictions of bipartitioning the qudits in the system \cite{enriquez2016maximally}.

In the spirit of these observations, it is also noteworthy that subsystem complementary recovery is not exact in quantum gravity, except in a simplifying limit that does not include gravitational interaction in the bulk.  
In fact, it has been proven in Ref.\ \cite{pollack2022understanding} that all holographic stabilizer codes do satisfy an operator-algebra version of complementary recovery \cite{harlow2017ryu}, which is more natural for quantum gravity systems. By using this more general sense of complementary recovery known as \emph{sub-algebra reconstruction}, it has been shown that holographic QRM codes do support transversal logical $T$ gates \cite{cao_qlego}. Even if we take a more conservative viewpoint, there is no reason a priori requiring a holographic code to be both a good quantum gravity toy model and a practical QEC code at the same time, if the design goal is ultimately fault tolerance. 

In this work, we delimit the restrictive condition of subsystem complementary recovery from holographic quantum code construction, favoring the sub-algebra reconstruction approach from Ref.\ \cite{pollack2022understanding}. Firstly, we review why the holographic QRM code does not obey subsystem complementary recovery, and that it does indeed possess a transversal logical $T$ gate. The erasure threshold for the asymptotically zero-rate version of the code is also calculated, and we find it to be well below the upper bound dictated by the zero-rate quantum erasure channel capacity. Recognizing (using the Choi-Jamiolkowski isomorphism) that holographic quantum codes can be viewed as a generalization of code concatenation on hyperbolic tilings, we leverage this relationship in order to construct several novel holographic codes which improve the erasure threshold of the holographic QRM code. Moreover, we observe that these new \emph{heterogeneous} holographic codes (\cref{fig:QRM.Steane.heterogeneous}) possess universal FT logical gate sets, allowing for the circumvention of the Eastin-Knill theorem. We additionally show that, by removing the central tensor(s) in our construction, one can construct a finite-rate ``black hole" tensor network code which allow for universal FT addressable gates. We then proceed to compare the erasure thresholds and physical space overhead of our codes to those of traditional ``tree-style" concatenated codes; the results indicate significant savings in space overhead with no sacrifice at the level of threshold performance. Our work therefore introduces a framework for constructing holographic quantum codes with universal FT logical gate sets, dispelling the myth of Clifford-group restrictions for FT logic in holographic stabilizer codes.

This paper is organized as follows: We begin by reviewing the formalism of stabilizer and subsystem stabilizer codes (\cref{section:stab_subsystemstab_codes_ft}), pieceable fault tolerance (\cref{section:pieceable_FT}), code concatenation (\cref{section:code_concatenation}), and holographic quantum codes (\cref{section:holographic_codes}). We then begin the results section (\cref{section:results}) by reviewing the holographic quantum Reed-Muller code, its lack of adherence to subsystem complementary recovery, its transversal implementations of several non-Clifford logical gates, and a short study of the threshold of the code under erasure (\cref{section:QRM_section}). Next, we show that holographic quantum codes can be viewed as a generalization of code concatenation for hyperbolic tilings (\cref{section:HQEC_code_concatenation}). Afterwards, we  demonstrate that heterogeneous holographic codes can be constructed which contain universal FT logical gates (\cref{section:universal_FT_logic}). \cref{section:finite.rate} describes the construction of our black hole tensor network code, and investigates its FT gate addressability properties. In \cref{section:thresholds_for_erasure,section:physical.growth.comparison} we describe how erasure thresholds for holographic codes can be analytically estimated; we then compare our estimations to numerical Monte Carlo results, and compare the scaling of physical qubits at the boundary to traditional tree-style concatenation. We provide concluding comments in \cref{section:discussion}. 

\section{Background} \label{section:background}

In this section, we introduce the mathematical and physical concepts needed to formulate the main results of later sections. In particular we review \emph{stabilizer and subsystem stabilizer codes} in \cref{section:stab_subsystemstab_codes_ft}; the relevant seed codes utilized in this work (\cref{section:seed.codes.review}); \emph{fault tolerance} (\cref{section:fault.tolerance}); the concept of \emph{pieceable fault tolerance} (\cref{section:pieceable_FT}); \emph{code concatenation} (\cref{section:code_concatenation}); and a brief review of \emph{holographic codes} in \cref{section:holographic_codes}.

\subsection{Stabilizer and subsystem stabilizer codes} \label{section:stab_subsystemstab_codes_ft}

Any code, classical or quantum, can be described by its \emph{code parameters} $\llbracket n,k,d \rrbracket$. In this work, $n$, $k$, and $d$ signify the number of physical qubits (or qudits) in the code; the number of encoded logical qubits which can be used to store and manipulate quantum bits; and the \emph{distance} of the code, which gives the minimal Pauli weight 
of any operator transforming one logical code word into another, respectively
\cite{lidar_qec_book,nielsen_chuang}. 

Broadly speaking, \emph{stabilizer codes} are by far currently the most common form of QEC code studied \cite{gottesman1997stabilizer,gottesman2016surviving}. Stabilizer codes are a class of quantum codes for which the \emph{code-space} $\mathscr{C}$ is defined as the simultaneous $+1$ eigenspace of Pauli stabilizer generators. More formally, the code-space is defined as
\begin{equation}
\mathscr{C} \coloneqq \{ \ket{\psi} \in \mathcal{H} : s\ket{\psi} = +\ket{\psi}, \forall s \in \mathcal{S} \}~,
\end{equation}
wherein the set of generators $\{ s \}$ which stabilize the code-space form an \emph{Abelian subgroup}, which is known as the \emph{stabilizer group} $\mathcal{S}$; as such, the set $\{ s \} \in \mathcal{S}$.

In addition to stabilizer generators, a Pauli stabilizer code is defined by its \emph{logical operators}, which are taken to be $\bar{X}$ and $\bar{Z}$, which together with the identity $\bar{I}$ and $\bar{Y}=\ii \bar{X} \bar{Z}$ form a basis for any other logical operator of the stabilizer code. We will follow the customary notation of writing any logical operator with an overline, e.g., $\bar{X}$ signifying a logical Pauli-$X$ gate.

The \emph{normalizer} $\mathcal{N}$ of a stabilizer group $\mathcal{S}$ defines the equivalence class of $n$-fold Pauli operations under which $\mathcal{S}$ is invariant. More succinctly, 
we define it as
\begin{equation}
\mathcal{N}(\mathcal{S}) \coloneqq  \{\mathcal{P} \in \mathcal{P}_{n}: \mathcal{P}\mathcal{S}\mathcal{P}^{\dagger} = \mathcal{S} \}~,
\end{equation}
where $\mathcal{P}$ is an element of the $n$-fold Pauli group $\mathcal{P}_{n}$. In this way, one can define the set of logical operators as $\mathcal{L} := \mathcal{N}(\mathcal{S}) / \mathcal{S}$, i.e., the quotient group of the normalizer by the stabilizer group itself \cite{lidar_qec_book}.

\emph{Subsystem stabilizer} codes are a subclass of the stabilizer category in which the code-space is further subdivided \cite{lidar_qec_book}. More precisely, the Hilbert space $\mathcal{H}$ is partitioned as 
\begin{equation}
\mathcal{H} = \mathcal{H}_{\mathscr{C}} \oplus \mathcal{H}_{\mathscr{\bar{C}}}~,
\end{equation}
where $\mathcal{H}_{\mathscr{C}}$ represents the Hilbert space of the code-space, and $\mathcal{H}_{\mathscr{\bar{C}}}$ represents its \emph{complement}; that is, everything that is not contained in the code-space. It is then possible to split up $\mathcal{H}_{\mathscr{C}}$ further as $\mathcal{H}_{\mathscr{C}} = \mathcal{H}_{\mathcal{L}} \otimes \mathcal{H}_{\mathcal{G}}$; here $\mathcal{H}_{\mathcal{L}}$ is the \emph{logical subspace} containing logical information, and $\mathcal{H}_{\mathcal{G}}$ is known as the \emph{gauge} 
(or junk) subspace, which contains extra degrees of freedom that can be used to measure stabilizers and diagnose syndrome information in the protected subspace $\mathcal{H}_{\mathcal{L}}$. 

The mechanism providing for such gauge degrees of freedom 
can be found by defining the gauge group $\mathcal{G}$ such that the stabilizer group $\mathcal{S}$ acts as the \emph{center} $\mathcal{Z}(\mathcal{G})$ of the group; that is, that 
\begin{equation}
\mathcal{Z}(\mathcal{G}) := \{iI_{n}, \mathcal{S}\} = \mathcal{C}(\mathcal{G}) \cap \mathcal{G}~.
\end{equation}
Here, $i{I}_{n}$ represents the $n$-fold identity operator with a phase, and $\mathcal{C}$ is the \emph{centralizer} of $\mathcal{G}$ such that 
\begin{equation}
\mathcal{C}(\mathcal{G}) := \{ \mathcal{O} \in \mathcal{G} : [\mathcal{O}, \mathcal{S}] = 0\}~,
\end{equation}
signifying that the centralizer is the set of operators inside the gauge group which commute with the stabilizer group. 

Often it can be useful to construct a max rank Abelian subgroup $\mathcal{S}'$ of the gauge group by complementing the stabilizer $\mathcal{S}$ with a mutually commuting subset $\mathcal{L}_G\subset \mathcal{G}$. This can be achieved, for example, by fixing the gauge qubits into particular stabilizer states and adding the logical identity operators that also stabilize the gauge qubits into $\mathcal{S}'$. Such a process is known as gauge fixing, and is often non-unique.

\subsection{Relevant seed codes}\label{section:seed.codes.review}

\begin{figure}
\centering
\includegraphics[width=\columnwidth]{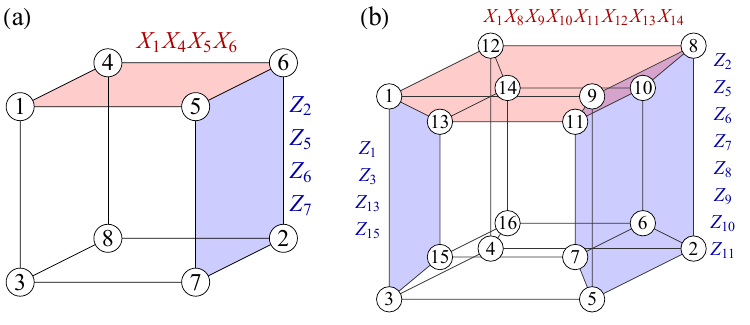}
\caption{Visualization of the stabilizers of the Steane and 
\emph{quantum Reed-Muller} (QRM) encoding tensor, following the format of Ref.\ \cite{raussendorf2012qec}. (a) The Steane encoding corresponds to an 8-qubit state that is stabilized by products of Pauli $X$ (red) and $Z$ (blue) operators acting on the qubits on the vertices of each of the six faces of the cube.
(b) The QRM encoding is stabilized by products of $X$ and $Z$ acting on the vertices of each of the seven cubes within the hypercube, in addition to products of $Z$ on the vertices of each of the twenty faces of the tiling. This overcomplete set of stabilizers can be reduced to generators corresponding to five $X$-cubes, five $Z$-cubes, and six $Z$-faces.
}
\label{fig:Steane-QRM-Stabs}
\end{figure}

We will use the following seed codes to assemble the holographic codes studied in this work.
A particularly simple class of stabilizer codes is given by \emph{Calderbank-Shor-Steane} (CSS) codes \cite{nielsen_chuang,steane1996multiple,calderbank1996good,lidar_qec_book}, in which the stabilizers and logical operators are each written purely in terms of products of either Pauli $X$ or Pauli $Z$ operators (plus identities).
A simple example is the 7-qubit \emph{Steane code} \cite{PhysRevLett.77.793}, which encodes one logical qubit and has also been the first CSS code applied in a holographic code \cite{Harris:2018jfl}.

One can write -- by invoking a simple isomorphism -- the 1-to-7 qubits encoding isometry (and its 
respective tensor representation) in terms of an 8-qubit state, whose stabilizers allow for a convenient visualization in terms of faces on a cube, shown in \cref{fig:Steane-QRM-Stabs}(a). 
To find the stabilizer for the code in which one (or more) of these vertices represents the logical qubit(s) to be encoded, one has to restrict oneself to the faces that do not touch the selected logical qubit(s).
For example, the Steane code for a single logical qubit (with index 1 in \cref{fig:Steane-QRM-Stabs}(a)) has the 7-qubit stabilizers
\begin{align}
\label{EQ_S_STEANE}
    \mathcal{S}_\text{Steane} = \langle & XIXIXIX, &&XXIIIXX, &&XIIXXXI, \nonumber\\
    &ZIZIZIZ, &&ZZIIIZZ, &&ZIIZZZI \rangle ~.
\end{align}
The logical operators are determined by taking the faces that do touch the logical qubit, leading to representations such as 
\begin{equation}
\bar{X}=IIXXXII 
\end{equation}
(red face in \cref{fig:Steane-QRM-Stabs}(a)) and 
\begin{equation}
\bar{Z}=IIZZZII.
\end{equation}
Note that the qubit 
ordering made use of here is somewhat non-standard in order to produce logical operator representations with support on blocks of nearby physical qubits, which will be helpful in later considerations.
We see that the Steane code manifestly belongs to a subclass of \emph{self-dual} CSS codes, in which both the set of stabilizers and logical operators is symmetrical under a Pauli $X \leftrightarrow Z$ exchange.
An example of a non-self-dual CSS code that we will use in this work is the 15-qubit \emph{quantum Reed-Muller} (QRM) code \cite{Steane:1996ci}, whose encoding tensor has stabilizers that can be visualized in terms of a hypercube, as shown in \cref{fig:Steane-QRM-Stabs}(b).
Similarly to the Steane code, the QRM code stabilizers for a single logical qubit are given by the cubes (for $X$ and $Z$) and inner faces (for $Z$) of this hypercube that do not touch the qubit with index 1, leading to
\begin{align}
\label{EQ_S_QRM}
    &\hspace{-5pt} \mathcal{S}_\text{QRM}= \nonumber\\ 
    \langle &XIXIXIXIXIXIXIX, &&XIIXXXXXXXIIIII,   \nonumber\\ 
    &XXXXXXIIIIIIIXX, &&IIIIXXIIXIIXXXX, \nonumber\\ 
    &ZIZIZIZIZIZIZIZ, &&ZIIZZZZZZZIIIII,   \nonumber\\ 
    &ZZZZZZIIIIIIIZZ, &&IIIIZZIIZIIZZZZ,  \nonumber\\ 
    &ZIZIIIZIIIZIIII, &&ZIIZIIZZIIIIIII, \nonumber\\ 
    &ZZZZIIIIIIIIIII, &&ZIIIZIZIZIIIIII ,\nonumber\\ 
    &ZIZIZIIIIIIIIIZ, &&ZIIZZZIIIIIIIII\rangle \ .
\end{align}
We can immediately see that the QRM code contains more pure-$Z$ than pure-$X$ stabilizers and is therefore not self-dual.
The logical operators for this code are similarly constructed, leading to e.g., $\bar{X}=IIIIIIXXXXXXXII$ (red cube in \cref{fig:Steane-QRM-Stabs}(b)) and $\bar{Z}=IZIIIIIIIIIZIZI$ (blue face).
The one non-CSS code relevant to our work is the $\llbracket 5,1,3 \rrbracket$ \emph{Laflamme code}, often simply referred to as the \emph{5-qubit code} \cite{Laflamme:1996iw}.
Its stabilizers, which are related to each other by a cyclic permutation of the physical qubits, are given by
\begin{equation}
\label{EQ_S_HAPPY}
\mathcal{S}_{513} = \langle XZZXI, IXZZX, XIXZZ, ZXIXZ \rangle~.
\end{equation}
Its logical operators can be written as $\bar{X}=XXXXX$ and $\bar{Z}=ZZZZZ$. As the 6-leg encoding isometry forms a \emph{perfect tensor}, it has also played a prominent role in the first proposal of a holographic quantum error correction code \cite{happy_code_paper}.

\subsection{Notions of fault tolerance} \label{section:fault.tolerance}

An important feature of a useful quantum code is \emph{transversality} of logical gates, defined as gates that do not spread errors between physical qubits. A transversal logical gate set is by design FT, as these logical gates do not couple the individual physical qubits of a code. 
Therefore, the introduction of one physical error cannot be spread to another part of the code block, avoiding a cascade of errors that could overwhelm the error correction scheme \cite{eastin_knill_theorem}. In this sense, one can consider transversal logical operations as naturally FT. For operations acting on a single logical qubit, transversal gates are simply those with a tensor product representation, e.g., the operator $\bar{X}$ of the $\llbracket 4,1,2 \rrbracket$ code from Ref.\ \cite{steinberg2024far}, which has representations $X_{1}X_{3}$ and $X_{2}X_{4}$ on the four physical qubits. For operations between multiple logical qubits (e.g., $\overline{CX}$), each assumed to be encoded by a separate copy (or \emph{code block}) of the quantum code, transversal gates may also couple a physical qubit in one code block with the same physical qubit in others.

Unfortunately, there are fundamental restrictions on the size of a transversal logical gate set. Essentially, no non-trivial quantum code (i.e., with $d \geq 2$) can admit a transversally implementable set of logical gates that is \emph{universal}, i.e., capable of approximating arbitrary logical unitaries. Instead, any such code will only be able to transversally perform gates in a discrete subgroup, such as the \emph{Clifford group}, a fundamental result known as the \emph{Eastin-Knill theorem} \cite{eastin_knill_theorem} (also studied in several other contexts, including \cite{yoshida_ft_logic,koenig_ftlogic}). 
In Ref.\ \cite{williamson_holographic_ft_logical_gates}, it has been shown that a class of \emph{holographic} codes which satisfy a condition known as \emph{complementary recovery} cannot admit transversal logical gate sets outside of the Clifford group either. In this work, we relax this condition and construct explicit examples of holographic codes which exhibit non-Clifford and universal transversal logical gate sets, as in 
Refs.\ \cite{hetero_tree1,hetero_tree2,hetero_tree3}. 
As there now exist several methods for 
in effect 
circumventing the Eastin-Knill theorem \cite{bravyi2005universal,bravyi2012magic,paetznick_code_switching,yoder_piece,gottesman_CX,anderson2014fault_conversion}, our work provides a further addition to this growing body of scientific literature. 

In order to allow for universal quantum computation, care must be taken in order to ensure that a given code possesses a method by which a circumvention to the Eastin-Knill theorem exists, and a universal FT logical gate set can be obtained. Some of the most commonly utilized \emph{universal gate sets} include $\{CX, T, H \}$ \cite{nielsen_chuang}, as well as $\{CCX, H\}$ \cite{shi2002both,aharonov2003simple}; even more generally, it has been shown in 
Ref.\ \cite{yoder_piece} that any representative from the set $\{CCX, CCY, CCZ\}$ and any single-qubit Clifford gate that does not commute with the former (e.g., the set $\{CCY, SHS\}$ is one example) is sufficient to guarantee universal quantum logic. As mentioned in \cref{section:intro}, various methods exist for circumventing the Eastin-Knill theorem and creating a universal FT logical gate set with a given quantum code; in this paper, we build on previous techniques developed in the context of code concatenation \cite{hetero_tree1,hetero_tree2,hetero_tree3,yoder_piece}.

\subsection{Pieceable fault tolerance} \label{section:pieceable_FT}

As a subclass of FT techniques, \emph{pieceable fault tolerance} (pFT) generally refers to a method of breaking down logical operators in their circuit representation into pieces which can be easily checked individually for the propagation of malicious physical errors throughout the circuit \cite{yoder_piece}. In between each piece of the logical operator's circuit, it is customary to perform an intermediate round of error correction with the portion of the stabilizer group which has remained constant. The overall evolution of the stabilizer group can be checked analytically for small codes in most cases, following the treatment of Ref.\ \cite{yoder_piece}. Many forms of pFT have been introduced over the years \cite{paetznick_code_switching,gottesman_CX,yoder_piece}, with the vast majority of research being invested into small atomic example codes. 

For the purposes of our work, we assume that all intermediate stages of syndrome extraction are ideal and therefore error-free. In particular, we focus on the construction of the $\overline{CCZ}$ gate from 
Ref.\ \cite{yoder_piece}, which is implementable for the $\llbracket 5,1,3 \rrbracket$ and Steane codes.  
\begin{figure*}
\centering
\includegraphics[width=\textwidth]{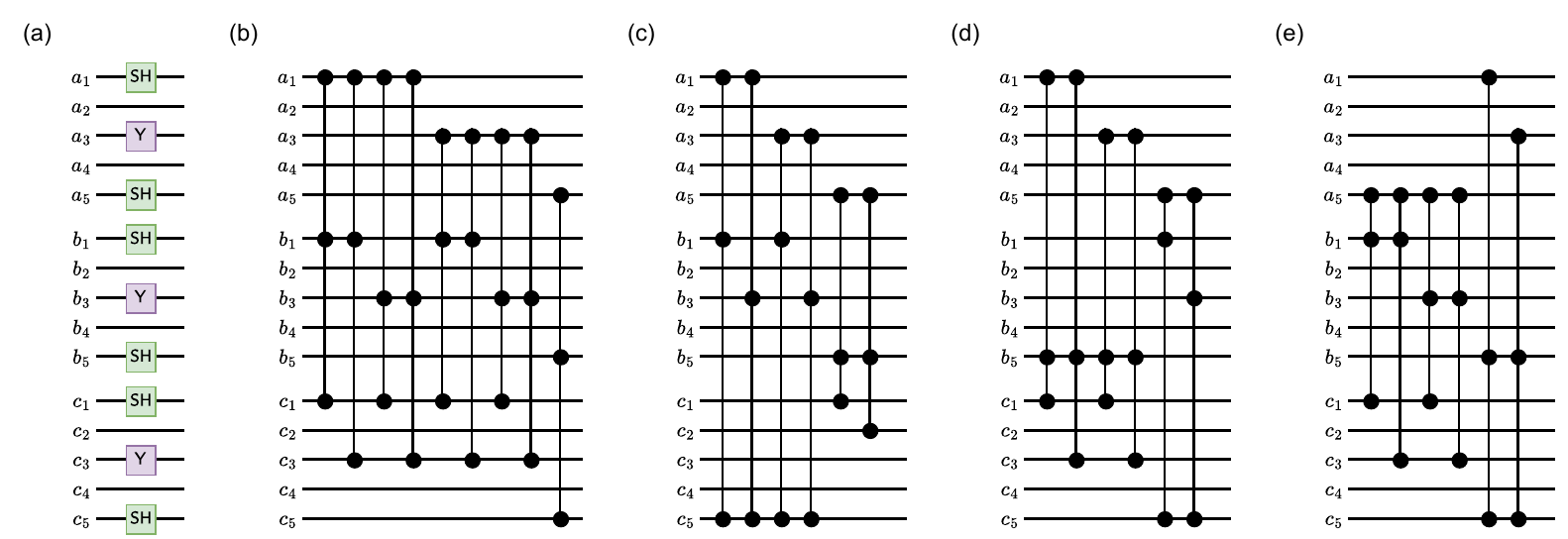}
\caption{pFT circuit-level implementation for the $\overline{CCZ}$ gate, figure adapted from Ref.\ \cite{yoder_piece}. We present its construction for the $\llbracket 5,1,3 \rrbracket$ code, wherein physical indices $\{a_{1}, a_{3}, a_{5}, b_{1}, b_{3}, b_{5}, c_{1}, c_{3}, c_{5}\}$ have been utilized from each of the three code blocks. For implementation in the Steane code \eqref{EQ_S_STEANE}, the circuit is exactly the same but has support on physical qubits $\{ a_{3}, a_{4}, a_{5}, b_{3}, b_{4}, b_{5}, c_{3}, c_{4}, c_{5} \}$ from all three code blocks. After each subcircuit is prepared, one round of intermediate error correction is performed, guaranteeing that the resulting circuit is 2-FT.}
\label{fig:pieceableCCZ}
\end{figure*}
The pFT construction for the $\overline{CCZ}$ gate is displayed in \cref{fig:pieceableCCZ}. We will give an overview of the procedure for the $\llbracket 5,1,3 \rrbracket$ code. As in \cref{fig:pieceableCCZ}(a), we first perform a sequence of $SH$ and $Y$ gates to physical qubits $\{a_{1}, a_{3}, a_{5}, b_{1}, b_{3}, b_{5}, c_{1}, c_{3}, c_{5}\}$ from all three code blocks. This transforms the $\llbracket 5,1,3 \rrbracket$ code's stabilizers $\mathcal{S}_{513}$ from \cref{EQ_S_HAPPY} and logical operators $\mathcal{L}_{513}$ into new groups $\mathcal{\tilde{S}}_{513}$ and $\mathcal{\tilde{L}}_{513}$
of the form
\begin{equation}
\mathcal{\tilde{S}}_{513} = \langle -YZXIZ, -ZZZXI, -IXZZZ, -ZIXZY \rangle~
\end{equation}
and 
\begin{equation}
\mathcal{\tilde{L}}_{513} = \langle ZIZIZ, XIXIX \rangle~.
\end{equation}
Next, we proceed by performing the circuit shown in (a). As has been discussed in Ref.\ \cite{yoder_piece}, each piece of the pFT circuit is 2-transversal, and as such transforms the logical operators of all three code blocks in a predictable manner. For all intermediate stages of error correction, the \texttt{PARSEC} method was utilized \cite{yoder_piece}. 

We summarize the \texttt{PARSEC} procedure as follows. As the circuit undergoes unitary evolution, the stabilizer group and logical operators also evolve, as discussed above. After each subcircuit from \cref{fig:pieceableCCZ} is performed, some stabilizers will still lie within the original stabilizer, while others will not; those that do remain we name \emph{constant}, and those that do not, we designate \emph{non-constant}. Following the treatment of Ref.\  \cite{yoder_piece}, we begin the \texttt{PARSEC} method by measuring all constant stabilizers from all code blocks. If two or more code blocks indicate an error, then we apply commensurate Pauli $X$ corrections. Conversely, if only one block indicates an error, then we measure the non-constant stabilizers next, and perform the appropriate correction. If no code blocks indicate an error, then we continue. More details on this method can be found in Ref.\ \cite{yoder_piece}. Here, 
we have provided a high-level survey of the scheme for the $\llbracket 5,1,3 \rrbracket$ code, but the process is identical for the Steane code as well, with only the caveat that physical qubits $\{ a_{5}, a_{6}, a_{7}, b_{5}, b_{6}, b_{7}, c_{5}, c_{6}, c_{7} \}$ are utilized instead. 

\subsection{Code concatenation} \label{section:code_concatenation}

A \emph{concatenated} (or \emph{tree tensor-network}, TTN) code is the conjoining of two or more codes in order to construct a larger one. In code concatenation, the physical output subspace for one code is used as the input logical subspace for the other. More rigorously, consider two codes: the first code, $\mathscr{C}_{1}$, possesses code parameters $\llbracket n_{1},k_{1},d_{1} \rrbracket_{\ell}$; another, $\mathscr{C}_{2}$, has parameters $\llbracket n_{2},k_{2},d_{2} \rrbracket_{\ell}$ \cite{gottesman2016surviving}. Here, $\ell$ refers to the local Hilbert space dimension for qudits. In our case, we set $\ell = 2$, which applies to both codes. The codes possess \emph{encoding maps} $\mathsf{W}_{1}, \mathsf{W}_{2}$, respectively, which can be defined as the set of gates that construct a logical code-space. That is, $\mathsf{W}_{1} \equiv \mathsf{W}(\mathscr{C}_{1})$.  

In order to perform code concatenation, the physical output subspace of $\mathscr{C}_{1}$ must be used as the logical input of $\mathscr{C}_{2}$. In this way, a new code with parameters $\llbracket n_{1}n_{2}, k_{1}, d_{\text{TTN}}\rrbracket$ and distance fulfilling $d_{\text{TTN}} \geq d_{1}d_{2}$ is formed; the encoding map for this new concatenated code takes on the form $\mathsf{W}_{2}^{\otimes n_{1}} \circ \mathsf{W}_{1}$. 

Code concatenation can be viewed in a \emph{tensor network} representation \cite{Ferris_2014}, by way of the \emph{quantum LEGO} (qLEGO) formalism \cite{farrelly2021tensor,cao_qlego}. qLEGO is a diagrammatic language by which one can combine smaller quantum error correction codes into larger ones, in effect generalizing code concatenation. We start by defining the quantum state vector $\ket{\psi}$ as
\begin{equation}
\label{eq:stateTN}
\ket{\psi} = T_{i_{1} ,\cdots,  i_{m}} \ket{i_{1} \cdots i_{m}}~,
\end{equation}
where the tensor $T_{i_{1} ,\cdots ,i_{m}}$ provides a specific decomposition of the vector state vector $\ket{\psi}$ in terms of the local basis $\ket{i_{1} ,\cdots, i_{m}}$. Furthermore, provided that $|\psi\rangle$ is sufficiently entangled, $T_{i_{1} ,\cdots, i_{m}}$ can be used in order to describe an \emph{isometry}, by specifying input and output legs of the tensor
\begin{equation}
\label{eq:isometryTN}
\mathsf{V}_{\ket{\psi}} := T_{i_{1} ,\cdots ,i_{m}} \ket{i_{k+1}, \cdots ,i_{m}} \bra{i_{1} ,\cdots ,i_{k}}~.
\end{equation} 
Here, $T_{i_{1} ,\cdots, i_{m}}$ represents a mapping from $k$ input legs to $n = (m-k)$ output legs.
One may readily form larger isometries by utilizing \emph{tensor contraction} techniques \cite{cao_qlego,farrelly2021tensor,farrelly2022local}. These combined isometries define the encoding maps for larger QEC codes; for the resulting code, one can always deduce the new stabilizer generators of the code by utilizing \emph{tensor pushing}. In general, the techniques discussed in the qLEGO formalism also apply to non-isometries, to non-Abelian stabilizer codes, as well as to non-additive codes \cite{shen2023quantum,kubischta2023family}. More details on qLEGO can be found at 
Refs.\ \cite{cao_qlego,fan2024lego_hqec}, but we will not make use of these more exotic applications in the current work.

\cref{fig:QRM.Steane.heterogeneous} shows some of the tensor-network code constructions that will be considered. 
\cref{fig:QRM.Steane.heterogeneous}(a) showcases the traditional tree-style concatenated code \cite{exact_concat_codes,knill2005quantum,gottesman1997stabilizer,gottesman2016surviving} for one logical qubit with a heterogeneous Steane/QRM structure, following the construction in 
Refs.\ \cite{hetero_tree1,hetero_tree2,hetero_tree3}. These codes are also expressible as \emph{tree tensor networks} (TTN), and for this reason we will use the shorthand \emph{TTN code} to refer to a tree-style concatenated code.
\cref{fig:QRM.Steane.heterogeneous}(b) and (c) display non-tree heterogeneous holographic codes for a single logical qubit. In (b) we show a Steane/QRM code with an initial Steane layer and in (c) a QRM/Steane code with an initial QRM layer. While we only draw the first three layers, each of these codes can be extended to arbitrarily many layers and physical qubits at asymptotically zero rate. 

\subsection{Holographic quantum error correcting codes} \label{section:holographic_codes}

Holographic tensor-network codes are a recently established class of subsystem stabilizer codes \cite{happy_code_paper,jahn2021topical,pollack2022understanding,harris_css}. In addition to their use as models of the \emph{Anti-de Sitter/conformal field theory} (AdS/CFT)  correspondence, holographic codes have an abundance of desirable properties for practical QEC: finite and customizable encoding rates; high thresholds under a wide variety of noise channels; structural versatility via the convenient qLEGO formalism \cite{cao_qlego}; distance scaling for logical qubits that is higher than $\sqrt{n}$, and conveniently-generated encoding circuits via \emph{graph-state} \cite{holography_encoding_graph} and \emph{ZX calculus} \cite{zx_encoding} methods. We display several of the holographic codes studied in this work in \cref{fig:extra.codes.and.BH.code,fig:QRM.Steane.heterogeneous}.

Holographic codes are constructed on \emph{uniform hyperbolic tilings} \cite{symmetriesofthings,adams2022tilingbook} whose symmetries mirror those of time-slices of AdS bulk vacua. The basic regular tilings that can be assembled from a single type of regular $p$-gon can be denoted by the \emph{Schl\"{a}fli symbols} $\{ p,q\}$, where $q$ stands for the number of tiles which meet at any vertex, and the tiling becoming hyperbolic for $\frac{1}{p} + \frac{1}{q} < \frac{1}{2}$. In the corresponding tensor network, we associate every vertex with a $q$-leg tensor, with a possible additional ``logical leg'' for the logical qubit to be encoded in $q$ physical ones.
However, the main focus of this work is on novel holographic codes on more general uniform hyperbolic tilings built from \emph{two} types of tiles, regular $q_1$-gons and $q_2$-gons, $p/2$ of each of which meet at every vertex (with $p$ necessarily even). We will usually consider the \emph{dual} of such alternating tilings, where vertices are bicolorable (corresponding to the tensors for two types of codes) and either $q_1$ or $q_2$ tiles surround each. These tilings are written as  $\{q_1,p/2,q_2\}$ in Schl\"{a}fli notation, or equivalently as $(q_1\; p/2\;  q_2)$ as the symmetries of the corresponding triangle group.
The condition for the tilings to be hyperbolic is $\frac{1}{q_1} + \frac{1}{q_2} + \frac{2}{p} < 1$, allowing for the possibility of infinitely many variations in such tilings.
In fact, this symmetry is slightly broken at the center for the zero-rate codes we consider, as only the central tensor has a central non-planar logical leg and hence a vertex with one fewer edge (see examples in \cref{fig:QRM.Steane.heterogeneous}).
As these tensor-network codes are built from two types of constituent codes (and corresponding \emph{seed tensors}), we name them \emph{heterogeneous} holographic quantum codes.

In contrast to TTN codes, wherein the encoding rate $r_{\text{TTN}}$ is limited by $k_{1}$, a holographic quantum code carries no such limitations because additional logical qubits can be injected from the outer codes during concatenation. In this work, we shall consider mostly constructions with asymptotically zero rate for simplicity. In spite of this limitation for this work, our arguments are easily generalizable to the finite-rate case, which we discuss in \cref{section:discussion}. 

\section{Results} \label{section:results}

\subsection{Holographic codes with non-Clifford transversal gates} \label{section:QRM_section}

It is relatively easy to build a holographic stabilizer code which does not obey subsystem complementary recovery and also possesses a non-Clifford transversal logical gate. The example that we begin with is the 15-qubit \emph{quantum Reed-Muller} (QRM) code with stabilizer group \eqref{EQ_S_QRM}, whose transversal logical gate set contains the gates $\{ \bar{T}, \overline{CX} \}$ in one gauge, and $\{ \overline{CCZ}\}$ in another \cite{paetznick_code_switching,cao_qlego}. This code is also related to the the family of \emph{higher-dimensional topological quantum codes} \cite{3d_color_code}. 
Furthermore, the QRM code has been used in conjunction with many other codes such as the Steane code in order to perform code switching \cite{anderson2014fault_conversion}, or to construct heterogeneous concatenated codes \cite{hetero_tree1,hetero_tree2,hetero_tree3}.

In the case of the holographic QRM code, one can utilize the tensor-pushing techniques of Ref.\ \cite{cao_qlego} to discern that the action of the transversal logical $\bar{T}$ gate from the central seed code onto the boundary physical qubits is transversal. Indeed, if we begin by labeling the central seed code as the zeroth layer, then every odd layer of \emph{edge inflation} of the quasi-periodic boundary of the tensor network \cite{jahn2022tensorqCFT,centralcharge_jahn,junyu_msc_thesis,conformalquasicrystals} necessarily requires application of $(T^{\dagger})^{\otimes n_{L}}$, where $n_{L}$ represents the total number of physical qubits at some layer $L$. For even layers, we simply apply $T^{\otimes n_{L}}$ instead. 

Let us examine why the holographic QRM code does not respect subsystem complementary recovery and therefore evades the no-go theorem of Ref.\ \cite{williamson_holographic_ft_logical_gates}. It is instructive to first examine a single $[[15,1,3]]$ QRM code. In the tetrahedral arrangement \cite{3d_color_code}, each minimal-weight $\bar{Z}$ operator is supported on 3 qubits on each edge of the tetrahedron whereas $\bar{X}$ operators are supported on a face of weight 7. As a simple example, consider a bipartition of the physical qubits into set $A$, which contains a single edge of the tetrahedron, and its complement $A^{c}$. It is clear that only the $\bar{Z}$ sub-algebra has support in either $A$ or $A^{c}$ whereas the $\bar{X}$ sub-algebra is unavailable to either subsystem (this is similar to certain bipartitions in the holographic Bacon-Shor and Evenbly codes, which utilize $\llbracket4,1,2\rrbracket$ codes \cite{steinberg2024far,cao2021approximateBScode}). Indeed, unless $A$ contains a single face, one does not have subsystem complementary recovery. This property is inherited by more general holographic QRM constructions when many QRM seeds are glued together \cite{cao_qlego}. One can show that, for a randomly chosen boundary subregion $A$, one can easily identify instances using the Gaussian elimination decoder \cite{harris_css,junyu_msc_thesis,fan2024lego_hqec} wherein the code sub-algebra on $A$ contains a non-trivial center because of the Abelian $\bar{Z}$ sub-algebra (but not the $\bar{X}$ sub-algebra) that one can recover on different bulk logical qubits. Nevertheless, as we mentioned earlier, the code is fully complementary in the sub-algebra sense \cite{pollack2022understanding}.

\begin{figure}
\centering
\includegraphics[width=\columnwidth]{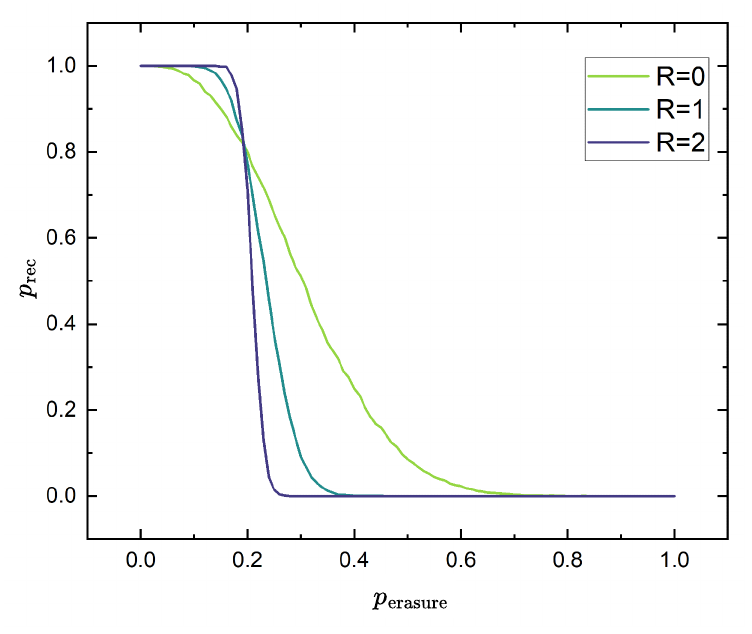}
\caption{Threshold for the asymptotically zero-rate holographic QRM code. A threshold of $19.2\%$ has been obtained using the Gaussian elimination decoder from Ref.\ \cite{fan2024lego_hqec}. 
Here, $p_{\text{erasure}}$ signifies the erasure probability of each physical qubit and $p_\text{rec}$ the recovery probability of the logical qubit. 
}
\label{fig:QRM_erasure}
\end{figure}

Analyzing the threshold of the holographic QRM code can be done using established methods, such as an optimal \emph{Gaussian elimination} decoder for the quantum erasure channel \cite{fan2024lego_hqec,harris_css,happy_code_paper}. We evaluate the erasure recovery threshold of an asymptotically zero-rate holographic QRM code; the results are shown in \cref{fig:QRM_erasure}. As is shown, the threshold of the code is $p_{\text{erasure}} \approx 19.2\%$. It is known that the quantum channel capacity for erasure is upper bounded at $50\%$ \cite{bennett1997capacities_erasure,wilde2013quantumIT}, compared to which the holographic QRM code performs poorly. 
Although the code admits a transversal non-Clifford gate, it does not include a universal gate set, lacking for example a transversal logical Hadamard $\bar{H}$; there thus exists no clear method to implement universal FT logic, in addition to its lackluster erasure threshold.

As we discussed in \cref{section:intro}, a heterogeneous concatenated code has been introduced in Refs.\ \cite{hetero_tree1,hetero_tree2,hetero_tree3}. There, the authors showed that universal FT logic is possible with this code, but that the FT threshold of the code is too low to permit practical feasibility. However, in the case of holographic codes, one can exploit a relationship between holographic and concatenated codes in order to constructed \emph{heterogeneous} holographic codes with universal FT logical gate sets; we will prove this relationship in the following section.

\subsection{Holographic quantum codes as generalized code concatenation} \label{section:HQEC_code_concatenation}

Let us begin by rephrasing concatenation in terms of a sequence of linear maps. Recall from \cref{section:code_concatenation} that the encoding map of a QEC code can be written as a linear isometry $\mathsf{V} :\mathcal{H}_{\mathcal{L}}\rightarrow \mathcal{H}_{\text{ph}}$ that maps from the logical Hilbert space to the physical Hilbert space. Since a concatenated code is one that takes the physical qubits of an inner code $\mathscr{C}_{1}$ and further encodes them as the logical qubit of an outer code $\mathscr{C}_{2}$, one can simply compose a sequence of isometric maps $\mathsf{V}_{\ell}, \mathsf{V}_{\ell-1},\dots, \mathsf{V}_0$  such that \footnote{For the sake of clarity, let $\mathsf{W}_{i}$ be the encoding isometry obtained from each seed code (or its shortened counterpart). We take $\mathsf{V}_{i}=\mathsf{W}_{i}\otimes I$ where $\mathsf{W}_{i}$ is the isometry induced by a single seed code/tensor and the identity operator $I$ acts on the remaining space where $\mathsf{W}_{i}$ does not have support.}
\begin{equation}
    \mathsf{V} =\mathsf{V}_{\ell}\circ \mathsf{V}_{\ell-1}\circ\dots\circ \mathsf{V}_0~.
\end{equation}
For each map $\mathsf{V}_{i}:\mathcal{H}_{L_{i}}\rightarrow \mathcal{H}_{\text{ph}_{i}}$, we have $\mathcal{H}_{\text{ph}_{i}}=\mathcal{H}_{\mathcal{L}_{i+1}}$  such that all the output qubits of $\mathsf{V}_{i}$ are encoded as logical qubits in the subsequent map $\mathsf{V}_{i+1}$; here $\mathcal{L}_{i}$ represents the $i^{\text{th}}$ level logical (input) qubits, and $\text{ph}_{i}$ are simply physical (output) qubits for the same level. When written as a tensor network, this is simply the contraction of isometric tensors where we compose an isometric map $\mathsf{V}_{i}$ each time we contract a tensor. 
For a more rigorous algebraic treatment of these inductive tensor network maps, consider \cite{chemissany2025infinite}.

Following this logic, holographic codes can be thought of as a generalized form of code concatenation. Here, we consider the concatenation through a mix of different codes. For homogeneous codes, we focus on a seed code $\llbracket n,1,d\rrbracket$ and its \emph{shortened} counterpart $\llbracket n-1,2,d-1\rrbracket$ \cite{gottesman1997stabilizer,raissi_modifying}.  For example, the zero-rate HaPPY code can be constructed simply as a concatenated code which encodes the physical qubits of the central seed code into $\llbracket 5,1,3\rrbracket$ and the shortened $\llbracket 4,2,2\rrbracket$ codes in the outer legs of each boundary-layer tensor; note that this concatenation strategy follows the individual tiles of a hyperbolic tessellation. In this work, we will refer to such shortened codes as \emph{child codes}. Technically, TTN codes can also be embedded onto regular hyperbolic tilings; these are known as the so-called \emph{ideal} $\{\infty,q\}$ tilings.

For the heterogeneous holographic codes featuring in this work, two different seeds $\llbracket n_1,1,d_1\rrbracket,\llbracket n_2,1,d_2\rrbracket$ and their child codes $\llbracket n_1-1,2,d_1-1\rrbracket, \llbracket n_2-1,2,d_2-1\rrbracket$ are used. For instance, a HaPPY-Steane heterogeneous holographic code can be defined by concatenating the central seed, a $\llbracket 5,1,3\rrbracket$ code, with five $\llbracket 7,1,3\rrbracket$ Steane codes in the next layer. Then these boundary qubits are further encoded into the shortened $\llbracket 4,2,2\rrbracket$ codes and $\llbracket 5,1,3\rrbracket$ codes in the subsequent layer. These qubits are then themselves encoded into a number of $\llbracket 7,1,3\rrbracket$ and the shortened $\llbracket 6,2,2\rrbracket$ codes in the following layer and so on.

A slight generalization is needed for the finite-rate codes, wherein logical qubits are also added along the way during concatenation. In this case, the homogeneous codes can be thought of as concatenating two types of shortened child codes $\llbracket n-1,2,d-1\rrbracket$, $\llbracket n-2,3,d-2\rrbracket$ together with the central seed code $\llbracket n,1,d\rrbracket$. However, instead of using all of the logical inputs of the seeds for encoding during concatenation, we leave some of them untouched. For example, in the finite-rate HaPPY code, the five qubits of the central seed code are encoded into the next layer using $\llbracket 4,2,2\rrbracket$ codes. However, we only use one logical input in the $\llbracket 4,2,2\rrbracket$ code during concatenation while leaving the other input free. The same follows for subsequent layers and the heterogeneous codes where, instead of concatenating one type of seed code and child code, we allow two or more atomic tensors in our construction. Mathematically, the encoding map $\mathsf{V}$ of such a code is still given by a sequence of isometric maps $\mathsf{V} = \Pi_i \mathsf{V}_{i}$ except the pre-image of $\mathsf{V}_{i+1}$ now strictly contain the range of $\mathsf{V}_{i}$. Therefore, this structure constitutes a natural generalization of code concatenation \cite{generalized_code_concat}. The error correction properties of holographic codes with many ungauged logical qubits are further discussed in \cite{Jahn:2025manylogical}.

Our finite-rate construction is also related to code concatenation in a different sense via the generalized \emph{channel-state duality} \cite{cao_qlego}. If we use this duality to designate all logical legs as physical legs in the holographic code except that of the central seed tensor, then each layer we are concatenating by a $\llbracket n,1,d\rrbracket$ or $\llbracket n-1,2,d-1\rrbracket$ child code. Consequently, not all of the physical qubits are encoded in the next layer of concatenation. In other words, in the dual description, the range of $\mathsf{V}_{i}$ can be strictly bigger than the domain of $\mathsf{V}_{i+1}$.

\subsection{Universal fault-tolerant logic in heterogeneous holographic codes} \label{section:universal_FT_logic}

Here, we describe three examples of \emph{heterogeneous} holographic codes which support universal FT gate sets. Our central strategy is the holographic concatenation of two seed codes with complementary gate sets, abstracting the approach for TTN codes proposed in Ref.\ \cite{hetero_tree1}. Additionally, we make use of the pFT formalism from Ref.\ \cite{yoder_piece}, generalized to the setting of holographic code concatenation.

\subsubsection{Holographic Steane/QRM code} \label{section:steane_QRM}
\begin{figure*}
\centering
\includegraphics[width=0.95\textwidth]{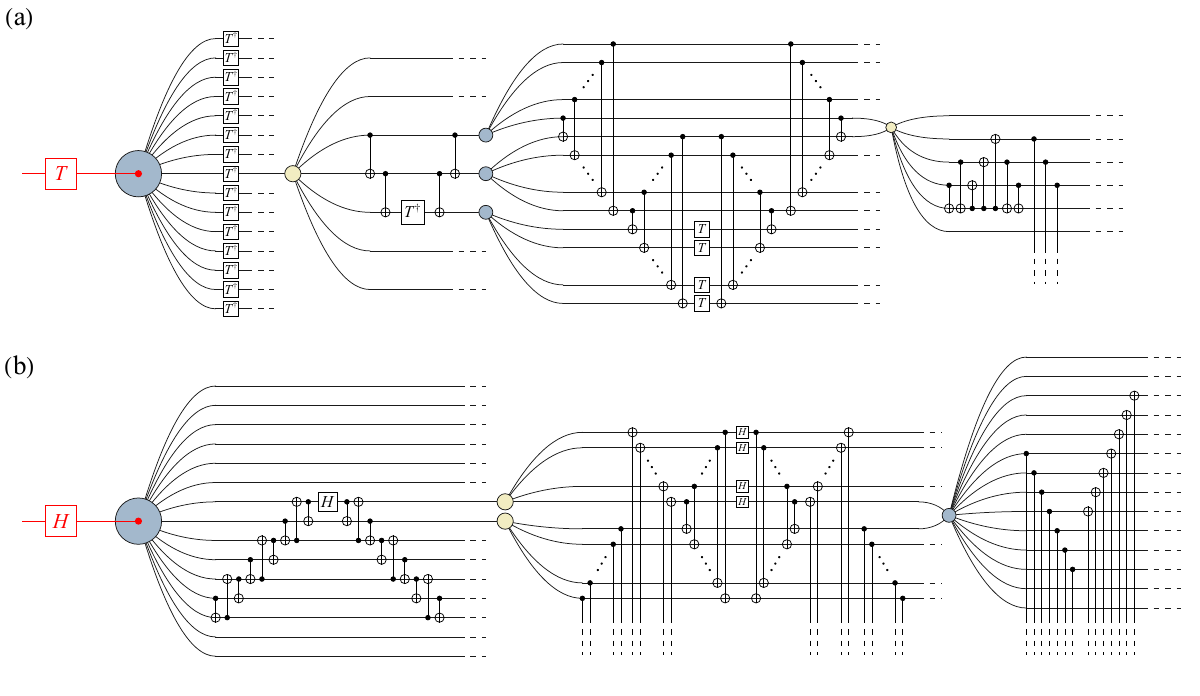}
\caption{Representation of logical operators (a) $\bar{T}$ and (b) $\bar{H}$ for the first few layers of the QRM/Steane code in \cref{fig:QRM.Steane.heterogeneous}(a). Circuits at every layer act equivalently on the code space (the final logical layer is highlighted in red), with the the last layer on the right only showing the representations of the first two gates of the previous layer. In this implementation, gates either act transversally or can be error-corrected into a transversal gate acting on the next layer. In contrast with TTN codes, branches reconnect after two layers, requiring tensors with two logical legs.}
\label{fig:pushing.TH.TN}
\end{figure*}
We begin by surveying a heterogeneous QRM/Steane code. Similar to the construction in Refs.\ \cite{hetero_tree1,hetero_tree2,hetero_tree3}, we begin by selecting either the QRM ($\mathscr{C}_{1}$) or Steane ($\mathscr{C}_{2}$) code as the central seed tensor for the zeroth layer of the holographic code. These variants are shown in \cref{fig:QRM.Steane.heterogeneous}. In \cref{fig:QRM.Steane.heterogeneous}(b), the Steane code $\mathscr{C}_{1}$ is taken as the central seed code, and is concatenated with the QRM code $\mathscr{C}_{2}$ in the subsequent layer; similarly, in (c) we exchange the central seed code for the QRM code, and concatenate in the second layer using the Steane code. Note that, for simplicity, we have assumed an asymptotically zero-rate version of the code. One can easily apply the methods from the previous section in order to create finite-rate constructions of the code; we treat an example of this in \cref{section:finite.rate}. 

We now show that this code admits a universal FT logical gate set. As stated in \cref{section:stab_subsystemstab_codes_ft}, a simple and commonly-referenced universal gate set for quantum computation consists of the gates $\{ CX, T, H \}$. In order to prove the existence of a universal FT logical gate set, one must demonstrate that all of the logical gates mentioned can be performed fault tolerantly. Other more general methods exist to prove fault tolerance for logical gate sets \cite{yoshida_ft_logic,williamson_holographic_ft_logical_gates,koenig_ftlogic}, but this method will serve our present purposes. As both the Steane and the QRM codes are CSS, Theorem 2.1 of Ref.\ \cite{cao_qlego} shows that any code conjoined from them must also be CSS. The $\overline{CX}$ gate is therefore transversal in this conjoined code.

The cases of $\bar{T}$ and $\bar{H}$ are more involved, however. Let us examine each logical gate separately. Very concretely, first consider the implementation of a logical $T$ gate in the code shown in \cref{fig:pushing.TH.TN}(a). A $\bar{T}$ on the central qubit is implemented transversally on the central QRM tensor as $(T^\dagger)^{\otimes 15}$. Each such $T^\dagger$ is then translated into a $\bar{T}^\dagger$ on the Steane code in the second layer, and can therefore be implemented in terms of (non-transversal) $CX$ and $T^\dagger$ gates on the third.
Repeating this procedure, we then map $\overline{CX}$ and $\bar{T}$ gates in the QRM code onto the fourth layer of the graph, where they again become transversal (in particular, with $\overline{CX}$ now acting transversally on different code blocks).
Note that we have defined our Steane and QRM codes such that non-transversal gates do not act on the first and last of the branching physical legs, ensuring that they are always fed into a code with $k=1$ input legs.
For the $\bar{H}$ gate, an analogous procedure unfolds, shown in \cref{fig:pushing.TH.TN}(b): We perform the Clifford circuit on the central QRM code, which consists of a single Hadamard and several $CX$. The Hadamard gate can be implemented transversally on the Steane layer, each of which can then be compiled into a physical Clifford circuit acting on the QRM tensors in the next layer. 
In summary, we can translate a non-transversal gate on one layer to a transversal one on the next layer. More detailed information on the implementation and tensor-pushing rules for $\bar{T}$ and $\bar{H}$ can be found in \cref{section:gate.prop.rules}.

On a conceptual level, error propagation in the holographic circuit shares many similarities with the proposal of Ref.\ \cite{hetero_tree1}, as the ``pushing'' of operators between layers is equivalent on seed codes with $k=1$ input legs, where the geometry appears locally tree-like. 
However, the uniform hyperbolic geometry of the holographic tensor network ensures that outgoing branches from any seed tensor reconnect after two layers, which leads to situations where operators act on $k=2$ logical input legs of a QRM or Steane child code. In \cref{fig:pushing.TH.TN}, this situation appears for $\bar{T}$ and $\bar{H}$ after four and three layers, respectively.
The breaking of naive transversality in these $k=2$ codes needs to be treated carefully, and one may worry that this could allow for a single physical error to spread to multiple logical ones uncontrollably, thus violating fault tolerance. 
For simplicity, let us focus on a QRM/Steane code with three layers (i.e., QRM-Steane-QRM) as in \cref{fig:QRM.Steane.heterogeneous}(c), which corresponds to the situation in \cref{fig:pushing.TH.TN} except that we ignore the last Steane layer of \cref{fig:pushing.TH.TN}(a).
Implementing a $\bar{T}$ gate on the central QRM code is then a transversal operation of $CX$ and $T$ gates acting on a total of 45 QRM code blocks on the last layer, thus posing to issues for fault tolerance: Applying error correction sequentially from the last to the first layer ensures that a single physical error remains correctable within each QRM block under application of a central $\bar{T}$, and is corrected before it reaches the second (Steane) layer.
For the logical Hadamard $\bar{H}$, the situation is more complicated: In addition to transversal $CX$ gates (sketched at the last layer of \cref{fig:pushing.TH.TN}(b)), we also find non-transversal $CX$ gates in each QRM code block on the last layer, arising both from $\bar{H}$ gates acting on any of these QRM codes as well as $\overline{CX}$ gates acting on $k=2$ QRM codes.
Careful inspection of the circuit in \cref{fig:pushing.TH.TN}(b) shows that a physical error on the last (QRM) layer can then propagate to a single error on up to four distinct code blocks in the Steane layer above (two for targeted Hadamard and four for targeted $\overline{CX}$).
However, since each block contains only one error, they remain correctable. 
More details on error propagation of targeted $\overline{CX}$ and $\bar{H}$ are shown in \cref{fig:CXpush} and \cref{fig:TandHpushing} in the Appendix.
As a result, all pairwise $\overline{CX}$, as well as a targeted $\bar{H}$, $\bar{T}$ can be implemented in a way that is tolerant against any single qubit error on the boundary with layer-by-layer error correction.
These errors will be corrected by performing syndrome measurement and error correction on the Steane layer. Hence the logical gates on the bulk qubit are FT. Then growing the code out to more layers allows it to be tolerant against higher-weight errors. What is more, one can remove the central QRM or Steane seed code to produce a ``black hole'' tensor network code, wherein one encodes 15 (or seven) logical qubits instead. We will treat this example of a finite-rate heterogeneous holographic code more meticulously in \cref{section:finite.rate}.

Taking stock, we can summarize the transversal logical steps for executing either $\bar{T}$ or $\bar{H}$ as the following: 
\begin{enumerate}
\setlength\itemsep{0pt}
\item Perform either of the circuits shown in \cref{fig:pushing.TH.TN}(a) or (b) in the QRM code blocks at the third layer,
\item execute one round of QRM-level quantum error correction, and
\item proceed with one round of Steane-level quantum error correction at the second layer. 
\end{enumerate}
We remark here that this scheme generalizes 
the techniques of Ref.\ \cite{hetero_tree1} to 
the realm of holographic code concatenation.

\subsubsection{Holographic HaPPY/QRM code} \label{section:happy_QRM}
\begin{figure*}
\centering
\includegraphics[width=0.95\textwidth]{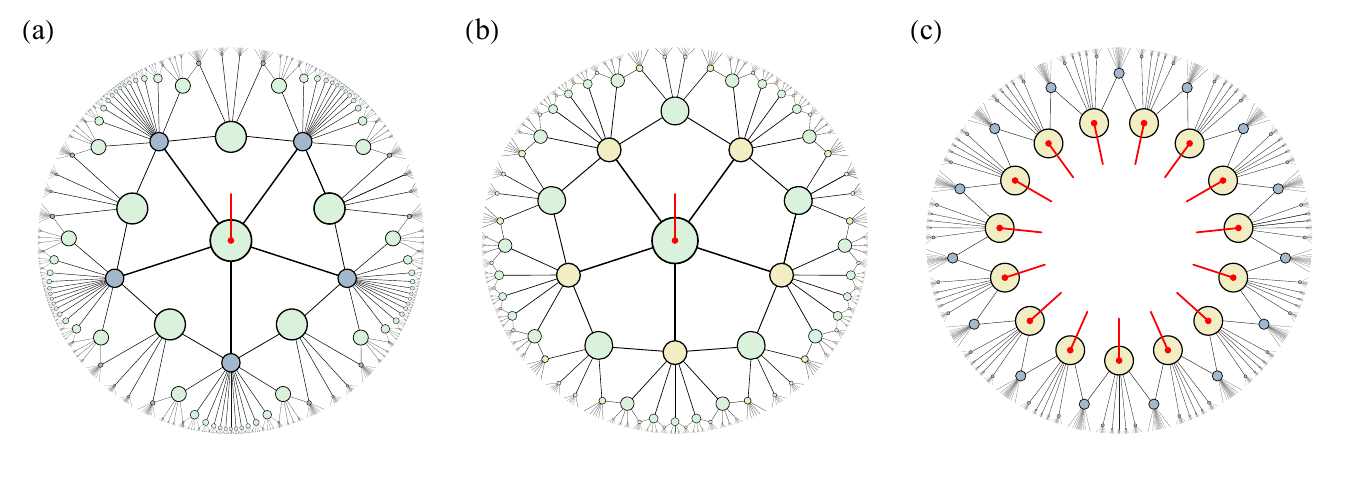}
\caption{Other heterogeneous holographic code constructions utilized in this work. (a) The alternating HaPPY/QRM code for one logical qubit (red tensor leg), defined on a $\{6,2,16\}$ hyperbolic tiling. QRM code tensors are shown in dark blue and HaPPY code tensors in light green. Each layer of tensors consists of entirely only QRM or HaPPY code tensors; this is true of all of the heterogeneous holographic codes that we have introduced in this work. The physical qubits of the code are identified with the black open legs on the tensor network boundary. (b) The alternating HaPPY/Steane code on a $\{6,2,8\}$ hyperbolic tiling. Steane code tensors are shown in light yellow. (c) A heterogeneous ``black hole code'' created by erasing the central tensor of a $\{16,2,8\}$ QRM/Steane code (\cref{fig:QRM.Steane.heterogeneous}(a)) and identifying the 15 internal open legs with as many logical qubits. For (a) and (b), one may alternatively choose to exchange the seed tensors for the $\mathscr{C}_{2}$ seed code instead, as in \cref{section:steane_QRM}.}
\label{fig:extra.codes.and.BH.code}
\end{figure*}
Alternatively, we may define code $\mathscr{C}_{1}$ as the $\llbracket 5,1,3 \rrbracket$ code, and $\mathscr{C}_{2}$ as the QRM code. As before, we can set either $\mathscr{C}_{1}$ or $\mathscr{C}_{2}$ as the inner code in the concatenation, which resides in the center of the hyperbolic tiling; for reference, we display the HaPPY/QRM heterogeneous code in \cref{fig:extra.codes.and.BH.code}. The $\llbracket 5,1,3 \rrbracket$ code allows for a pFT circuit realization of $\overline{CCZ}$ (depicted in \cref{fig:pieceableCCZ}), and also contains a transversal $\overline{SH}$ gate. In a complementary way, the QRM code contains a transversal $\overline{CCZ}$ \cite{paetznick_code_switching,3d_color_code}, the usual $\bar{T}$, and a FT circuit implementation of $\bar{H}$, as has been  described in \cref{fig:pushing.TH.TN}(b). Using these ingredients, it is possible to create a universal logical gate set for this heterogeneous holographic code; that is, our universal FT logical gate set is constructed from $\{CCZ, SH\}$. As has been stated in 
Ref.\ \cite{yoder_piece}, one can indeed always utilize a single-qubit Clifford gate that does not commute with one from the set of $\langle CCX, CCY, CCZ\rangle$ in order to complete a universal FT gate set. 

Implementation of the $\llbracket 5,1,3 \rrbracket$ code's $\overline{CCZ}$ is accomplished following the piecewise procedure described in \cref{section:pieceable_FT}. As described in that section, in between each of the four steps we require intermediate error correction rounds by means of the \texttt{PARSEC} formalism described in Ref.\ \cite{yoder_piece}. As described, individual errors can accumulate and spread throughout each circuit piece. However, we reiterate that each piece of the $\overline{CCZ}$ circuit is known to be 2-FT; this means that at most two errors will spread throughout the code block during each piece's execution. Subsequently, one round of ideal error correction ensures that errors will not spread further, and that at most one physical error will propagate to the next piece of the circuit structure. As the tensor representing the $\llbracket 5,1,3 \rrbracket$ corresponds to an \emph{absolutely maximally-entangled} (AME) state \cite{enriquez2016maximally,ame_states1}, we are allowed to utilize the innermost physical indices at each $\llbracket 5,1,3 \rrbracket$ level, thereby avoiding the spread of errors into child codes of the QRM code at subsequent layers. 

The final ingredient required is the compilation of the $\overline{SH}$ gate for the QRM layers. Just as in \cref{section:steane_QRM}, one can perform the circuit for $\bar{H}$ from \cref{fig:pushing.TH.TN}(b). The only caveat now is that one must additionally perform two applications of transversal $\bar{T}$, in order to push $\overline{SH}$ to the $\llbracket 5,1,3 \rrbracket$ layer of the code. 

\subsubsection{Holographic HaPPY/Steane code} \label{section:happy_steane}

Having established universal FT logic for the previous two example heterogeneous holographic codes, our final example of the HaPPY/Steane holographic concatenation is straightforward. Firstly, both the $\llbracket 5,1,3 \rrbracket$ and Steane codes admit pFT circuit-level implementations of the $\overline{CCZ}$ gate; the only differences lie in the physical qubits involved in the pFT circuit and the transformations of the stabilizers. As we have already shown that it is possible to perform the required logical gates by acting only on the middle wires of the tensors representing the Steane and $\llbracket 5,1,3 \rrbracket$ codes because they are dual to PME and AME states respectively, we can apply the same ideas for this heterogeneous code as well, thus ensuring FT for every stage of the pFT circuit of the $\overline{CCZ}$ gate. 

In order to complete the logical gate set, we utilize the transversal $\overline{SH}$ gate from the $\llbracket 5,1,3 \rrbracket$ code, as in \cref{section:happy_QRM}. Here, application of the $\overline{SH}$ gate is easy in the Steane layer, as both $\bar{S}$ and $\bar{H}$ are transversal for the Steane code. Therefore, we have proven that the heterogeneous holographic HaPPY/Steane code also exhibits universal FT logic.

Note that the above pFT protocol enables FT universal logic on homogeneous holographic codes like the HaPPY code or the holographic Steane code as well. One may then ask why it would be advantageous at all to consider a heterogeneous HaPPY/Steane holographic code instead. There are two separate reasons for doing so. Firstly, the proposed HaPPY/Steane code utilizes less space overhead than the homogeneous holographic Steane code; this is evident by the construction of the code. Secondly, Appendix D of Ref.\ \cite{yoder_piece} presented a two-stage, 21-gate implementation of the $\overline{CCZ}$ gate for the Steane code, in lieu of the 27-gate, four-stage implementation required for the $\llbracket 5,1,3\rrbracket$ code; this variant would allow for large time and space overhead savings as the concatenated code is grown layerwise.

\subsubsection{Finite-rate heterogeneous holographic codes} \label{section:finite.rate}

The methods that we utilized in \cref{section:steane_QRM,section:happy_QRM,section:happy_steane} in order to ensure universal FT logic for asymptotically zero-rate heterogeneous holographic codes translate well to the regime of finite-rate holographic codes as well. As an example, consider the scenario of the QRM/Steane code, which encodes into the central logical qubit, either from a QRM code or a Steane code.. Now, in order to form finite-rate versions of this code, we remove the central logical qubit, and utilize the logical qubits from the subsequent layer directly as an encoding block; this is shown in \cref{fig:extra.codes.and.BH.code}(c), in which we have created a ``black hole" in the bulk of the tensor network. This scenario, originally proposed in 
Ref.\ \cite{happy_code_paper}, allows us to encode 15 logical qubits, with the distance of said logical qubits being determined by the number of layers of holographic concatenation before reaching the boundary (physical) qubits. 

Going further, one may be able to construct an entire class of such ``black hole'' codes by cutting out black holes of larger radii in the center of the bulk to encode more logical qubits while growing the tensor network layerwise at the physical level. In our interpretation, the black hole is shown to ``grow" layerwise with the number of physical qubits, and allows for more logical qubits to be encoded on its ``horizon''. Depending on the distance and overall rate of the code's logical qubits that are desired, the number of layerwise holographic concatenations separating the physical qubits from the logical layer will determine both of these code parameters. In the language of continuum AdS/CFT, one can say that the bulk IR cutoff and the temperature of the black hole together fix the code parameters. If we allow for the number of layers in between the black hole and boundary to increase, then we obtain an asymptotically zero-rate code whose distance increases in the usual way for holographic codes; otherwise, if we grow the black hole layerwise as we increase the number of physical qubits, we obtain an asymptotically finite-rate code with constant distance.

One may ask that, as we have added many logical qubits into the bulk, whether or not the logical gates of individual logical qubits remain addressable. For the case of the QRM/Steane code we can answer in the affirmative, subject to some additional conditions involving the reflection symmetries of the Steane code; these additional considerations are treated in \cref{section:black.hole.code.appendix}. 

\subsection{Thresholds for erasure}\label{section:thresholds_for_erasure}

Heterogeneity also has potential benefits for raising error thresholds. 
Leveraging the relationship explained in \cref{section:HQEC_code_concatenation} allows us to directly relate holographic codes to code concatenation, permitting the exact calculation of many QEC-related properties, such as thresholds under diverse noise channels and quantum weight enumerators, inter alia \cite{cao2023quantum,cao2024expansionpack,exact_concat_codes}. As a demonstration, we calculate thresholds with respect to \emph{qubit erasure}; this is carried out analytically, using the treatment from Ref.\ \cite{exact_concat_codes}, and numerically, utilizing the recent software tool \texttt{LEGO\_HQEC} \cite{fan2024lego_hqec}.  Here, we define the \emph{quantum erasure} (or \emph{qubit-loss}) channel as 
\begin{equation}
\mathcal{E}_{e}(\rho) = (1-\epsilon)\rho + \epsilon \text{Tr}[\rho]\ket{e}\bra{e}~,
\end{equation}
wherein the channel erases a qubit with probability $\epsilon$ and does nothing with probability $(1-\epsilon)$ \cite{wilde2013quantumIT,bennett1997capacities_erasure}.

There are several reasons for considering erasure errors. Firstly, the erasure channel itself has been proposed as a benchmark for basic resilience of a quantum error correction code \cite{delfosse2016linear,Connolly_2024,Delfosse_2020}; as a result, erasure thresholds are simple and yet powerful coding-theoretic indicators for the code properties at capacity level. Secondly, erasure errors or located errors constitute an important class of errors that are highly relevant for practical error correction. For example, qubit loss is ubiquitous in quantum systems, particularly in those experiencing photon loss, such as in quantum communications setups \cite{erasure.stace}. Recent experiments have also shown the usefulness of converting biased-noise Pauli noise or leakage error into erasure channels
by carefully monitoring individual qubits in Rydberg-atom, trapped-ion, and superconducting quantum devices \cite{wu2022erasure_conversion,scholl2023erasure,kang2023quantum,Chou:2023kol}. 

Let us start by reviewing a simple example of naive concatenation. Consider a seed code that encodes a single logical qubit and is subjected to \emph{independently-and-identically-distributed noise} (iid) in the form of a single-qubit erasure channel of erasure probability $p$. Let $F(p)$ be the logical  error probability of erasure where we count the logical information as erased if there exists any logical operator that is not accessible from the remaining non-erased qubits. The function $F(p)$ is easy to obtain for a small seed code, either through analytic calculation or through Monte Carlo sampling. Now consider a naive concatenation of the same seed in a TTN code with $\ell$ layers. Suppose that the outermost seed code has logical error $F(p)$, then $p_{\rm inner}=F(p)$ is the physical error rate for the immediately adjacent inner layer code. The logical erasure error probability of the encoded qubit at the center is then given by nesting the same function $\ell$ times $p_{L}=F(F(\dots F(p)))$.

The logical erasure error probability can be generated by understanding how the physical erasure errors on the boundary qubits get transformed as they move along a graph geodesic which connects the boundary site(s) to the center of the bulk. Each seed code we encounter along the path carries its own distinct error suppression function $F_i(p)$ where $i=1$ denotes the outermost layer, $i=2$ the immediately adjacent inner layer adjacent and so on.  Because the logical error rate of the outer seed code transforms into the physical error rate of the inner seed code, traversing this path produces precisely the sequence of nested functions such that the final logical erasure probability $p_L$ is given by $p_L=F_{\ell}(F_{\ell-1}(\dots F_1(p))$. In fact, by repeatedly nesting similar functions in a multilayer concatenation scheme, one can also determine the effective logical error channels \cite{exact_concat_codes}. In the above example, since all $F_i(p)=F(p)$, it is possible to deduce the erasure threshold from the property of just a single seed code. For any non-trivial seed code that can correct erasures, $F(p)<p$ as long as $p<p_{\rm th}$ where $p_{\rm th}=F(p_{\rm th})$. 

Recall that zero-rate holographic codes can be viewed as generalized concatenated codes that possess a wider variety of concatenation patterns to choose from. For example, instead of encoding each physical qubit into a distinct code block, scenarios now emerge wherein two adjacent qubits are encoded into the same code block. By examining which seed codes one passes through whilst traversing similar boundary-to-bulk geodesics, the same argument as that from above can be repeated in order to estimate the erasure threshold. Thanks to the quasi-periodic nature of holographic codes, geodesics originating from different boundary sites no longer pass through the same sequence of seed codes. Indeed, UV-to-IR propagation can now pass through two distinct tensor configurations generated by the seed code and its child with error functions $F_{s}(p)$ and $F_{c}(p)$, respectively. Furthermore, the order and frequency at which one encounters these different error functions are also path-dependent, even for a homogeneous holographic code \cite{cao2021hyper}. In the thermodynamic limit, for a path chosen at random, these distinct segments of $F_{s}(p)$ and $F_c(p)$ can appear with some fixed probability. Therefore, we approximate holographic concatenation as a probabilistic concatenation. 
The error rate calculation then simplifies considerably by homogenizing the different error paths (i.e., geodesics) using a weighted sum of the different seed codes and contraction patterns it may encounter such that for each layer $i$, $F_i(p)\mapsto F(p;q_s,q_c)=q_{s} F_s(p)+q_c F_c(p)$ where $q_{s}$ and $q_c$ are the probabilities of encountering a seed or its child in layer $i$ and $q_s+q_c=1$. Since the holographic tensor network is approximately scale-invariant in the thermodynamic limit, one can take these probabilities to be independent of the layer number.

While the precise probability distribution of $\{q_{i}\}$ depend entirely on the tiling (and thus network connectivity), obtaining these values can be technically involved. Instead, we provide upper and lower bounds on the logical error rate by skewing the distribution towards the $F_i(p)$ which represent the best- or worst-case seed codes for correcting erasure. From these weighted functions $F_i(p;q_s,q_c)$, we can then obtain the lower- and upper-bound limits of threshold estimates in a manner similar to the TTN code calculation.
\begin{figure*}
\centering
\includegraphics[width=\textwidth]{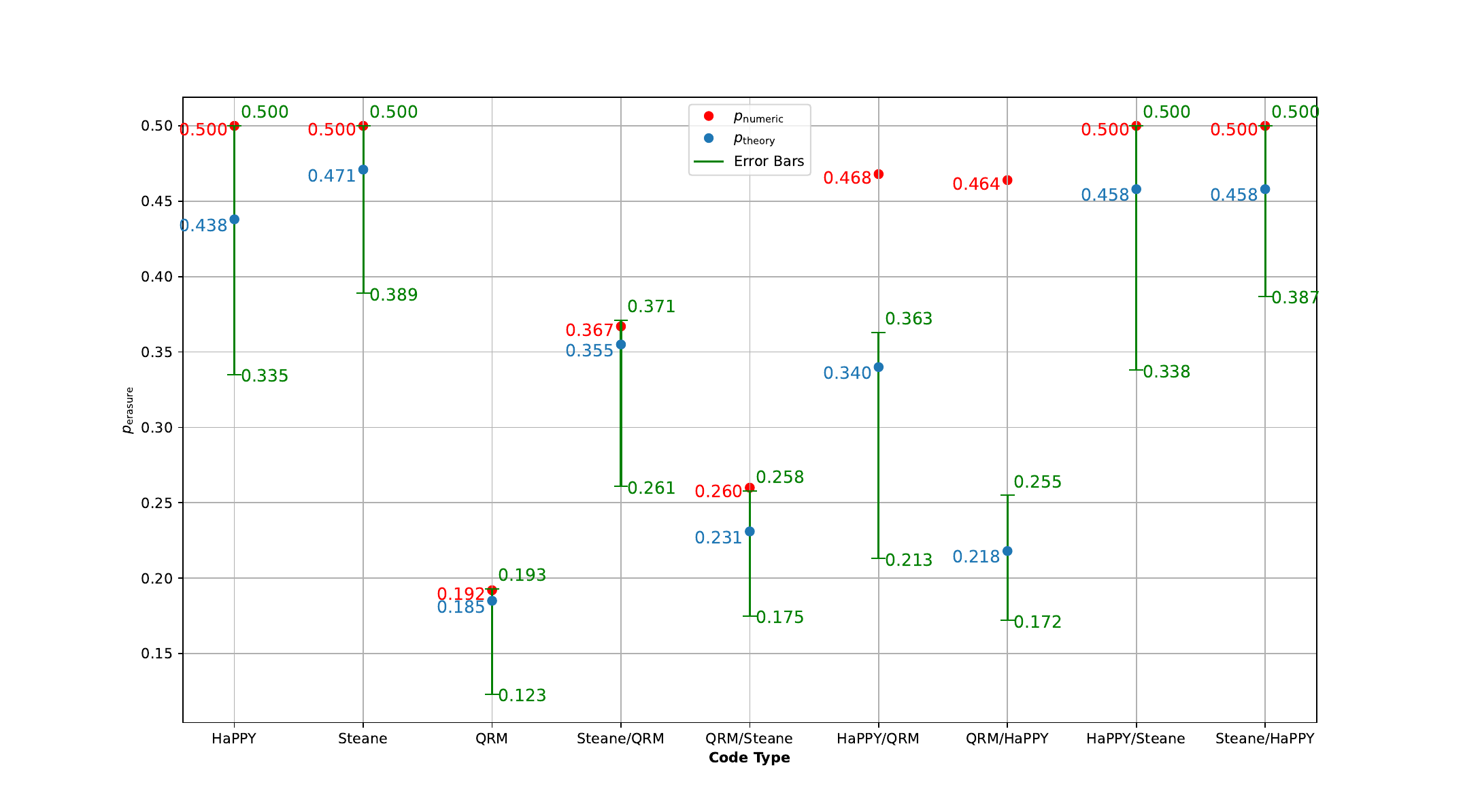}
\caption{From left to right we display the thresholds $p_{\text{erasure}}$, as obtained from numerics ($p_{\text{numeric}}$), estimated from our theoretical analysis ($p_{\text{theory}}$), as well as for upper and lower bounds (shown in green), without considering sub-algebra reconstruction. These thresholds pertain to the following asymptotically zero-rate holographic codes: HaPPY; holographic Steane; holographic QRM; heterogeneous Steane/QRM; heterogeneous QRM/Steane; heterogeneous HaPPY/QRM; heterogeneous QRM/HaPPY; heterogeneous HaPPY/Steane; and heterogeneous Steane/HaPPY. As is evidenced, our analytical estimate provides a useful lower-bound estimate of erasure threshold performance, as compared with the actual numerical value. The upper and lower bounds relate whether or not we assume worst-case and best-case traversing through the bulk; that is, either having errors pass solely through $d=3$ codes, or solely through error-detecting $d=2$ codes. We remark also that both the heterogeneous code constructions utilizing the $\llbracket 5,1,3 \rrbracket$ code either come close to or attain quantum channel capacity for an asymptotically zero-rate instance. We also remark here that the TTN codes from \cref{section:code_concatenation} obtain erasure thresholds represented by the green upper bound lines in our plot.}
\label{fig:thresholds.codes}
\end{figure*}
A similar process holds for heterogeneous holographic codes; the main difference is that the seed codes and their children now double in our examples. For clarity, denote the two weighted error functions as $G_a(p)$, $G_b(p)$, corresponding to the two distinct seed codes.  The error path also encounters distinct types of seed codes and children in a particular ordering. However, one key difference in the heterogeneous case lies in differences that depend on the layer which we terminate the concatenation on. If we take two distinct codes $\mathscr{C}_{1}$ and $\mathscr{C}_{2}$ for our heterogeneous holographic code, then choosing to terminate the concatenation on an odd layer necessarily ends up with code $\mathscr{C}_{2}$ at the boundary (in addition to its children); conversely, if we terminate the concatenation at even layers, then $\mathscr{C}_{1}$ and its children are expected to appear closest to the boundary. We stress also that this pattern applies only to edge inflation of hyperbolic tilings, with vertex inflation leading to combinations of both code types at any layer \cite{jahn2022tensorqCFT,centralcharge_jahn,conformalquasicrystals,junyu_msc_thesis}. 

Furthermore, when estimating error thresholds via our probabilistic error-path method, one always passes through the two types of seed codes (or their children) in a particular order, due to the layering pattern, thereby reducing to two differing compositions of the effective error functions $F_{ba}(p)=G_a(G_b(p))$ and  $F_{ab}(p)=G_b(G_a(p))$. For each of the two distinct outermost layer configurations, an estimate of the threshold can be extracted the same way as before.  Generally, $F_{ab}\ne F_{ba}$ and the type of code that the boundary layer terminates in plays a significant role in deciding the erasure threshold, as shown by the numerics as well as the analytical estimates (\cref{fig:thresholds.codes}). In particular, we see that the threshold structure bifurcates, as having a better code against erasure on the outmost layer always produces a higher erasure threshold; indeed similar results were found for TTN codes in \cite{chamberland2017complementarygauge}. Intuitively, having an outer layer that corrects erasure errors better than the inner layer is more effective than the reverse simply because this scheme lowers the erasure probability below the threshold of the inner-layer code, whereas the opposite arrangement often raises the error probability above the inner-layer threshold. 

In \cref{fig:thresholds.codes}, we report the recovery thresholds for erasure ($p_{\text{erasure}}$) according to numerical Monte Carlo simulations executed using \texttt{LEGO\_HQEC} \cite{fan2024lego_hqec} ($p_{\text{numeric}}$), estimations from our theoretical analysis ($p_{\text{theory}}$) above, and upper- and lower-bound estimations (in green), taken without factoring in sub-algebra reconstruction. As is apparent from the figure, all of the theoretical and numerical estimates fall squarely within the upper and lower bounds dictated by our analysis. In this way, our theory-level analyses prove to be a reliable and 
tight lower bound to the numerical results obtained. The only exception to our empirical rule can be found in the HaPPY/QRM heterogeneous holographic code, for which we found numeric thresholds that far outstripped even our upper-bound estimates. This is likely because of the undercounting of erasure recovery because the analytical bound considers a qubit erased if any of its sub-algebra is erased. This is generally too stringent. As a final note, we have included all of the estimations utilized in our study in \cref{section:analysis.thresholds}.

As a remark, we also note some important consequences of heterogeneous holographic codes. Firstly, the holographic codes built from the same seed codes can exhibit thresholds far higher than those of the TTN counterparts. For example, the heterogeneous TTN codes built out of $[[5,1,3]]$ and QRM codes have erasure thresholds $0.363$ and $0.255$ depending on the order of concatenation, whereas the holographic counterparts have $0.468$ and $0.464$, respectively, in the same layer ordering. Furthermore, heterogeneous doping in holographic codes can be used to raise erasure thresholds of homogeneous codes. For example, a homogeneous holographic QRM code has threshold $0.192$ whereas a Steane/QRM code has threshold $0.367$ instead.

\subsection{Growth of physical system size}\label{section:physical.growth.comparison}

Compared to heterogeneous TTN codes \cite{hetero_tree1}, holographic codes can also provide substantial saving in space overhead. As discussed above, the geometry of heterogeneous holographic codes is given by an alternating hyperbolic $\{q_1,p/2,q_2\}$ tiling, with tensors placed on each vertex and all connected edges corresponding to contractions. To build such a tensor network iteratively, we add alternating layers of both types of tensors following \emph{edge inflation}, i.e., adding a single tensor to every open leg.
Due to the hyperbolic geometry, the number of open boundary legs and hence physical sites of the code increases exponentially with the number of layers. We now compute this growth exactly. For each type of tensor labeled by $k=1,2$, we can define two types of tensors at the last layer of the tensor network: Tensors $a_k$ that are connected to the remaining tensor network via only one leg and tensors $b_k$ that are connected via two legs. A single edge inflation step then corresponds to the replacement rule
\begin{align}
    a_{1,2} &\mapsto (a_{2,1})^{q_{1,2}-3}\, b_{2,1} \ , \\ 
    b_{1,2} &\mapsto (a_{2,1})^{q_{1,2}-4}\, b_{2,1} \ .
\end{align}
Here powers denote repetitions of letters, e.g., $(a_1)^3 = a_1 a_1 a_1$ or three consecutive tensors of type $a_1$. Note that even though each tensor is connected to two $b_{1,2}$ tensors on the next layer of an inflation step, each of these $b_{1,2}$ tensors connects two previous branches and thus only appears once in the inflation rule (whether to include it at the beginning or the end of the replacement rule is a matter of convention).

After each step, the type of the boundary tensors switches between the two. In one double step, the above inflation rule then acts on the first type as
\begin{align}
    a_1 &\mapsto \left( (a_1)^{q_2-3} b_1 \right)^{q_1-3} (a_1)^{q_2-4}\, b_1  \ , \\ 
    b_1 &\mapsto \left( (a_1)^{q_2-3} b_1 \right)^{q_1-4} (a_1)^{q_2-4}\, b_1 \ .
\end{align}
Counting the number of each tensor type as $n_{a_1}$ and $n_{b_1}$, this double-inflation thus acts on the vector $\vec{n}=(n_{a_1},n_{b_1})^\text{T}$ as a matrix multiplication $\vec{n} \mapsto M \vec{n}$ with the \emph{substitution matrix}
\begin{equation}
    M = \begin{pmatrix}
    (q_1-2) (q_2-3) - 1 & (q_1-3)(q_2-3)-1 \\
    q_1-2 & q_1-3
    \end{pmatrix} \ ,
\end{equation}
The asymptotic scaling factor in each double-step is given by the larger of the two eigenvalues of this matrix, which are
\begin{equation}
    \lambda_{1,2}(q_1,q_2) = \frac{x-2 \pm \sqrt{x(x-4)}}{2} \ , \; x=(q_1-2)(q_2-2) \ .
\end{equation}
The number of physical qubits of the resulting code is then the number of open, uncontracted legs on the boundary tensors,
\begin{align}
    n_\text{phys} = n_{a_1} (q_1-1) + n_{a_2} (q_1-2) \ .
\end{align}
Note that this differs from the number of tensor created in the next layer, as these may connect to more than one of the open legs.

As an explicit example, consider the QRM/Steane code with $q_1=16$ and $q_2 = 8$. The central QRM code tensor, which we denote by another letter $o$, branches into 15 legs ending in a Steane code tensor, or as a double-inflation step,
\begin{equation}
    o \mapsto (a_2)^{15} \mapsto ((a_1)^5 b_1 )^{15} \ ,
\end{equation}
with $\vec{n}^{(1)}=(75,15)^\text{T}$. 

After a total of $i$ double-inflation steps, we then find for the number of boundary vertices the expression
\begin{align}
    \vec{n}^{(i)} &= M^{i-1} \vec{n}^{(1)} \nonumber\\
    &= 
    \begin{pmatrix}
        \frac{3 (35-3 \sqrt{105})}{14} \\
        \frac{3(-10+\sqrt{105})}{4}
    \end{pmatrix}
    (\lambda_1)^i
    + 
    \begin{pmatrix}
        \frac{3(35+3 \sqrt{105})}{14} \\
        -\frac{3(10+\sqrt{105})}{4}
    \end{pmatrix}
    (\lambda_2)^i \ .
\end{align}
for $\lambda_{1,2} = 41 \pm 4 \sqrt{105}$.
The number of physical qubits after $i$ double-steps, all emanating from a last layer of QRM-code tensors, is then given by
\begin{align}
    n_\text{phys}^{(i)} &= 15 n_{a_1}^{(i)} + 14 n_{b_1}^{(i)}\\
    &= \frac{3 (35+ 4\sqrt{105})}{14} (\lambda_1)^i
    +\frac{3 (35- 4\sqrt{105})}{14} (\lambda_2)^i \ .\nonumber
\end{align}
For the first few steps, this result evaluates to $n_\text{phys}^{(0)}=15$, $n_\text{phys}^{(1)}=1335$ and $n_\text{phys}^{(3)}=109455$, increasing by a factor of around $\lambda_1 \approx 82$ every step.

In order to resolve intermediate layers, 
e.g., applying the double-inflation step with the last layer composed of Steane-code tensors, we can simply use the above results and exchange $q_1 \leftrightarrow q_2$ to form a new substitution matrix $M^\prime=\left.M\right|_{q_1 \leftrightarrow q_2}$. We then begin inflation with a vector $\vec{n}^\prime = (n_{a_2},n_{b_2})=(15,0)$ and apply ${M^\prime}^{j-1} \vec{n}^\prime$, leading to
\begin{align}
    {n_\text{phys}^\prime}^{(j)} &= 7 n_{a_2}^{(i)} + 6 n_{b_2}^{(i)}\\
    &= \frac{735+72\sqrt{105}}{14} (\lambda_1)^i
    +\frac{735-72\sqrt{105}}{14} (\lambda_2)^i \ .\nonumber
\end{align}
Explicitly, ${n_\text{phys}^\prime}^{(0)}=105$, ${n_\text{phys}^\prime}^{(1)}=8625$, and ${n_\text{phys}^\prime}^{(2)}=707145$.
This growth in the number of physical sites with each layers can be further increased by choosing a larger loop length $p$, i.e., reconnecting fewer branches in the tensor network. In the limit $p \to\infty$, this results in a \emph{tree tensor network} (TTN) code with a particularly simple inflation rule: Every even odd and even layer increases the number of physical sites by a factor of $q_1-1$ and $q_2-2$, respectively. In the QRM/Steane example, we then find $n_\text{phys}^{(i)}=105^i$ with $n_\text{phys}^{(2)}=11025$ and $n_\text{phys}^{(3)}=1157625$, somewhat faster than for the $p=4$ case. 

For a large number of layers, we thus find that heterogeneous holographic codes require exponentially fewer physical qubits than TTN codes, while still obtaining similar thresholds (the only exception to this is the HaPPY/QRM code, which substantially outperforms the TTN code's threshold, as shown in \cref{fig:thresholds.codes}). For the QRM/Steane seed tensors discussed here, both the holographic and TTN code require $n^{(1)}=105$ physical qubits after one double-layer, while after two we already find a reduction of $1 - 8625/11025 \approx 21.8\%$ in the number of physical qubits.
Considering single-layer inflation, e.g., a three-layer QRM-Steane-QRM code, this amounts to a space overhead saving of $10\sim 30\%$ with $O(10^2)\sim O(10^4)$ qubits even just at three to five layers.

\section{Discussion} \label{section:discussion}

In this work, we have taken steps to bring holographic quantum error correcting codes substantially closer to practical considerations in quantum computing. While for a number of years, such codes have been studied solely because of their conceptually interesting properties, we here show that they are actually a lot more plausible from a practically minded perspective as has been anticipated. Specifically, we have
shown how to construct heterogeneous holographic codes whose \emph{fault-tolerant}  (FT) logical quantum gate sets circumvent the Eastin-Knill theorem. We have proven that these schemes are possible by viewing the structure of holographic codes as a generalization of traditional tree-style code concatenation to the setting of hyperbolic tilings. Universal FT logic has been  established for three examples of heterogeneous holographic codes, as all codes admit FT universal logical gate sets via either complementary logical gate sets (in the case of the holographic QRM/Steane code), or by utilizing logical gate sets in tandem with pFT circuits (as in the holographic HaPPY/QRM and HaPPY/Steane codes). 

Putting this connection between practical quantum error correction and holographic codes upside down,
moreover, we have provided an example of a finite-rate heterogeneous holographic code inspired by the study of black holes in AdS space. We have analyzed and demonstrated that this new subclass of holographic codes not only exhibits the same rate and distance tunability that typical holographic codes possess, but crucially also provides a resolution to issues of bulk logical qubits possessing different distances, as well as targeted logical gate addressability thereof.

Taking advantage of the relationship to concatenated codes has permitted us to estimate the threshold of various holographic codes against erasure; we compared these estimates to results obtained via numerical simulations, largely finding agreement of numerical results within upper-bound ranges given using our estimation method. Compared to homogeneous holographic codes and/or heterogeneous TTN counterparts, heterogeneous holographic codes can substantially raise the erasure threshold and reduce space overhead, a tendency that increases as the size of the code grows layerwise. Our results not only refute the findings of Ref.\ \cite{williamson_holographic_ft_logical_gates} (i.e., that holographic codes can only possess transversal logical gate sets within the Clifford group) in an intricate way, but also exemplify how to construct high-threshold and finite-rate holographic quantum codes equipped with universal FT logic. All of these results are summarized in \cref{table:summary}. 

\begin{table*}
\begin{tabular}{ |c|c|c|c|c|c| } 
\hline
Code type & Universal logical gate set & $p_{\text{theory}}$ & $p_{\text{erasure}}$ & $p_{\text{upper}}$ & $n(L=0,1,2,3)$ \\
\hline
Steane/QRM & $\{\bar{H}, \bar{T}, \overline{CX}\}$ & 0.355 & 0.367 & 0.371 & 7, 105, 679, 8617 \\ 
QRM/Steane & $\{\bar{H}, \bar{T}, \overline{CX}\}$ & 0.231 & 0.26 & 0.258 & 15, 105, 1335, 8625 \\ 
HaPPY/QRM & $\{\overline{SH}, \overline{CCZ}\}$ & 0.34 & 0.468 & 0.363 & 5, 75, 345, 4055 \\ 
QRM/HaPPY & $\{\overline{SH}, \overline{CCZ}\}$ & 0.218 & 0.464 & 0.255 & 15, 75, 885, 4065 \\ 
HaPPY/Steane & $\{\overline{SH}, \overline{CCZ}\}$ & 0.458 & 0.5 & 0.5 & 5, 35, 145, 775\\ 
Steane/HaPPY & $\{\overline{SH}, \overline{CCZ}\}$ & 0.458 & 0.5 & 0.5 & 7, 35, 189, 777 \\ 
\hline
\end{tabular}
\caption{Table summarizing the principle contributions of our work. From left to right, columns denote the code type utilized, its universal logical gate set, the erasure thresholds $p_{\text{theory}}$, $p_{\text{erasure}}$, and the threshold of the related tree-style concatenated code, as estimated by the upper bound $p_{\text{upper}}$, and the number of physical qubits on the boundary at the first few layers $L$ of concatenation, respectively.}
\label{table:summary}
\end{table*}

To complete this summary, a 
few comments are still in order. In \cref{section:thresholds_for_erasure}, we have observed that the numerical threshold vastly outperformed the analytical threshold estimates if a non-CSS code was paired with a CSS code. A major contribution to this discrepancy is likely due to the definition of an erasure error in a seed. Thus far we have been working with subsystem erasures where a logical qubit is counted as erased if any of its logical operators are erased. However, a sizable contribution of this erasure does not erase the entire logical qubit, but rather just a logical sub-algebra, such as a $\bar{X}$ or $\bar{Z}$. This is especially apparent in QRM codes where only a sub-algebra is supported on a small subsystem, in stark contrast with the perfect code where having access to a subsystem either allows access to the entire encoded qubit or none of it. Therefore, erasures in HaPPY/QRM codes are dominated by sub-algebra erasures, which are counted as stronger subsystem erasures in the more conservative analytical estimations. However, generally sub-algebra erasures do not lead to subsystem erasures in a concatenated scheme. For example, consider a perfect 5-qubit code at the center of the bulk which admits weight-3 representation of logical operators $\bar{X}=IZXZI, \bar{Y}=ZIYIZ$. Suppose erasure on the outer layers forbids us from reconstructing the $X$ operators on qubits 1, 2, 4, and 5, one can still recover the full logical Pauli algebra. As such, the sub-algebra contribution can significantly raise the threshold in these setups. On this note, we remark that it is impressive that mixing non-CSS with CSS seed codes in our setup can produce much better thresholds than expected. In any case, we also note that the numerical thresholds calculated for our HaPPY/QRM code approaches the zero-rate \emph{quantum channel capacity} for the erasure channel, while for the HaPPY/Steane code, it achieves the bound calculated in Ref.\ \cite{bennett1997capacities_erasure}. This fact implies that such a heterogeneous holographic code could be a prime candidate for potential experimental implementations, although it is not yet known how heterogeneous holographic codes perform under more general noise channels, such as depolarizing or biased noise \cite{fan2024overcoming}. We leave more a more detailed treatment of these questions for future work.

The finite-rate holographic ``black hole'' code that we have discussed in \cref{section:finite.rate} has found to have many useful features: tunable rate and distance parameters, as well as addressable gates due to the relatively low degree of causal cone overlap from encoded logical qubits to the boundary. Other proposals for building finite-rate holographic codes have been discussed, yielding asymptotic variants \cite{farrelly2021tensor,happy_code_paper,harris_css,harris_int_opt,farrelly2022parallel,steinberg2024far}; however, one issue for these constructions is the fact that the distance for logical qubits in the bulk, although better protected than in TTN codes \cite{alex_logical_recovery}, still differs, depending on a given logical qubit's relative distance to the physical boundary. Another issue concomitant to the previous matter involved \emph{gate addressability}, as placing logical qubits in the bulk with overlapping geodesic support may prove troublesome when considering logical quantum computation in the bulk. Our proposal averts both issues by promoting entire layers of logical qubits, while maintaining that the physical support of all logical qubits does not overlap too greatly. The cost of this procedure involves either trading off distance scaling, or the asymptotically finite-rate property. Some of the rate/distance trade-off can be further improved by permitting a changing curvature going from the boundary to the bulk such that a less hyperbolic bulk interior geometry compared to the asymptotic boundary will increase encoding rate without sacrificing distance. 

As other code classes (notably \emph{quantum low-density parity-check codes} \cite{breuckmann2021quantum}) also have suffered both from gate addressability concerns and issues regarding universal FT logic, our result shows the promise of holographic quantum codes in the practical regime. One may even envision a scenario in which \emph{gauge fixing} techniques (as in \cite{steinberg2024far}) could be used on lower-distance logical qubits in the bulk to ease the requirements for fault tolerance in heterogeneous holographic codes; we leave such exploration for future work.

It has  recently been shown in 
Ref.\ \cite{steinberg2024far,steinberg_2023} that a derivative of the \emph{hyperinvariant tensor networks} of Refs.\ \cite{steinberg2022conformal,evenbly.hyperinvariant}, called \emph{Evenbly codes}, greatly benefit from applying \emph{gauge-fixing} techniques in order to alter the code's error correction and FT logical-gate properties. As the QRM code itself typifies a 2-uniform quantum many-body state, it stands to reason that one may be able to construct homogeneous or heterogeneous QRM versions of the Evenbly code. We leave this idea for future work as well.

Finally, as our method provides the first instances of universal FT logic in holographic quantum codes, one can ask whether it is possible to tune the logical gate set in seed codes in order to optimize the fidelity of a particular gate decomposition for a target unitary \cite{sarkar2024yaqq}. Indeed, \cite{kubischta2023family} has experimented with constructing codes with more exotic transversal logical gate sets, in addition to novel code switching techniques which have recently emerged \cite{ouyang2024measurement}.  Our method of constructing resilient codes with universal FT logical gate sets may be adaptable to this setting as well. Furthermore, \cite{cao_lackey_targeted} suggests that by applying quantum Lego formalism, it is possible to build up holographic codes that support addressable transversal/unit-depth multi-qubit (non-Clifford) gates which are FT within a single code block by choosing the seed tensors carefully. All in all, our method can be extended to even larger numbers of seed tensors, giving rise to more complex hyperbolic tiling patterns; such a design space allows for the construction of holographic quantum codes with arbitrary FT logical gate sets, depending on compilation constraints in distinct quantum computer architectures. Generalizing this idea into a full-fledged framework will be the subject of upcoming research. 

\section*{Acknowledgments}

\subsection*{People}

We would like to thank Aritra Sarkar for insights regarding universal gate sets and the concept of \emph{weakly} universal gate sets \cite{aharonov2003simple,shi2002both}, and Joseph Sullivan for mentioning that the gate set $\{CCZ, SH\}$ is also an example of a universal quantum gate set, as well as Michael Perlin and Sivaprasad Omanakuttan for useful discussions.  

\subsection*{Funding}

MS and SF thank the Intel Corporation for financial support. AJ and JE have been supported by the Einstein Foundation Berlin. JE has additionally been supported by the DFG (CRC 183), the BMBF (QSolid, RealistiQ, PasQuops), the Munich Quantum Valley, the Quantum Flagship (Millenion, PasQuans2), and the European Research Council (DebuQC).

\subsection*{Software resources}

The following library has been utilized in this project: \texttt{LEGO\_HQEC} \cite{fan2024lego_hqec}. 

\subsection*{Author contributions}

JF conducted the numerical simulations, utilizing the software package \cite{fan2024lego_hqec} as part of his master thesis, under the supervision of MS and SF. MS, AJ, and CC systematically proved the existence of universal FT logic for all three example heterogeneous holographic codes. CC developed the weighted-approximation method for analyzing thresholds under erasure for all of the holographic and TTN codes featured. MS and CC proved the relationship between code concatenation and holographic codes. AJ generated the hyperbolic tiling figures, and performed the space-overhead calculations from the discussion. MS, AJ, and CC wrote the paper. JE and SF supervised the work and guided the general project direction alongside CC and AJ.  

\subsection*{Disclaimer}

This work has been prepared for informational purposes by the Global Technology Applied Research center of JPMorgan Chase \& Co. This paper is not a product of the Research Department of JPMorgan Chase \& Co. or its affiliates. Neither JPMorgan Chase \& Co. nor any of its affiliates makes any explicit or implied representation or warranty and none of them accept any liability in connection with this paper, including, without limitation, with respect to the completeness, accuracy, or reliability of the information contained herein and the potential legal, compliance, tax, or accounting effects hereof. This document is not intended as investment research or investment advice, or as a recommendation, offer, or solicitation for the purchase or sale of any security, financial instrument, financial product or service, or to be used in any way for evaluating the merits of participating in
any transaction.

\clearpage
\bibliography{bibliography}

\clearpage
\appendix 

\section{Gate addressability details for the black hole tensor network code}\label{section:black.hole.code.appendix}

Regarding the QRM/Steane construction, 
we discuss first how to implement targeted gates on the 15 encoded qubits feeding into the central QRM tensor, which has been replaced with a black hole, in the present scenario. We will refer to this set of qubits as $\mathcal{Q}$. First we discuss targeted $\overline{CX}$ gates on any pair of logical qubits. Note that although the code has a global transversal $\overline{CX}$ because it is CSS, which is sufficient for producing a FT $\overline{CX}$ on the central bulk qubit, we actually have far more control in the holographic black hole code. 
\begin{figure*}
\centering
\includegraphics[width=0.9\linewidth]{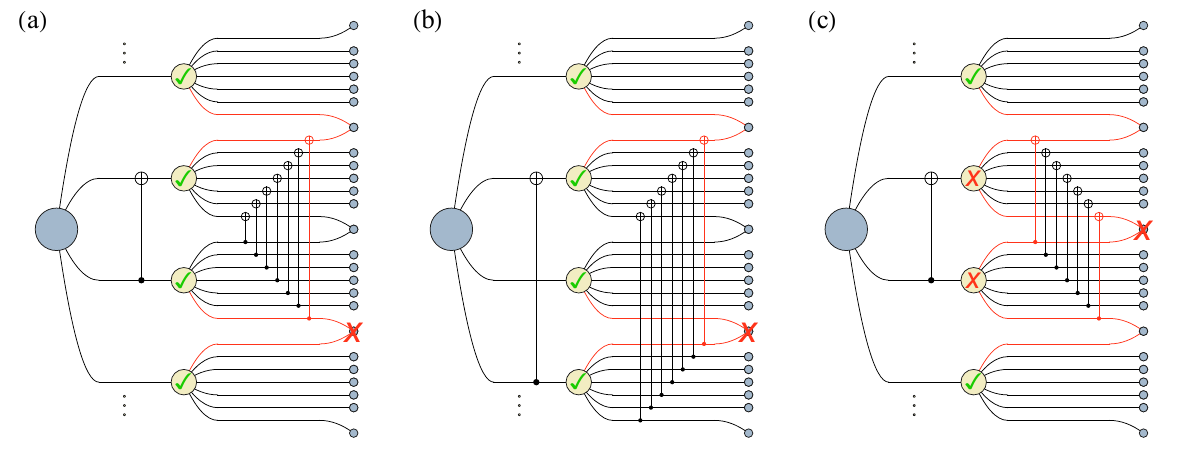}
\caption{Fault tolerance dependence on qubit ordering in the QRM/Steane code.
(a) A single $\overline{CX}$ gate applied to two Steane code blocks (bright yellow disks) can be expressed as seven $CX$ gates on the physical layer. In this ``reflected'' ordering, a single error on one of the subsequent QRM code blocks (dark blue disks) will only propagate to a single, correctable error within each of the four affected Steane blocks when applying $\overline{CX}$, preserving fault tolerance.
(b) For $\overline{CX}$ acting on non-adjacent Steane code blocks, a regular ordering of physical qubits also preserves fault tolerance.
(c) A regular ordering for adjacent Steane code blocks can propagate a single QRM error to two errors each in two Steane blocks, leading to a non-FT $\overline{CX}$ gate. However, if we also make use of the location information from the outer layer, then the special arrangement in (a) may not be necessary. As the only error that can propagate inwards without being caught by error correction must be located on the outermost edges (red) of each tensor in (c), two such located errors can also be corrected by a $d=3$ code.
}
\label{fig:CXpush}
\end{figure*}
Any $\overline{CX}$ acting on adjacent qubits in $\mathcal{Q}$ will push to transversal $CX$ on the Steane layer. They can then be compiled as a 2-local Clifford circuit on the outermost QRM layer that have support in the boundary subregion  (\cref{fig:CXpush}a). The $CX$s that are pushed to the $k=1$ QRM tensors on layer two can be implemented transversally. The $CX$s that are pushed through the $k=2$ QRM child codes will be compiled differently on the boundary, but remain as a finite depth 2-local Clifford circuit.

We see that if any single error occurs on the part of the boundary qubits feeding into the $k=1$ QRMs, any one error will spread to two distinct $[[15,1,3]]$ QRM codes. Although the total weight of the error has grown to two, each QRM block receives a single error and hence they are correctable. If any error occurs on the QRM child code with $k=2$, then it can spread to four logical errors on the outermost QRM layer (\cref{fig:CXpush}a). However, we again emphasize that, since each of the four errors propagates to a distinct Steane block on the next layer, they can also be corrected, ensuring fault tolerance. 

It is important that the transversal $CX$ gates on the Steane layer can be implemented this way instead of the usual bitwise transversal $CX$ where identical qubits on two Steane codes are connected by a $CX$. We see this is possible thanks to a permutation symmetry present in the Steane code.

Consider \cref{fig:Steane-QRM-Stabs}(a): Fixing the logical qubit (vertex 1) leaves a rotational symmetry of the cube along the axis from vertex 1 to 2. This manifests itself in an invariance of the code under the permutation $\sigma=(2)(3,4,5)(6,7,8)$. 
Hence a $\overline{CX}$ can be obtained by either applying $CX$s across identical qubits on both blocks (that is, as $0-0,1-1,2-2,\dots$, with the first physical index labeled as 0), or by applying CXs on a permuted set of qubits (i.e., $0-0, 1-2, 2-3,3-1,4-6,6-4,5-5$). Such a condition is required, as the non-reflected implementation of the transversal $CX$ can cause errors to propagate out of control, as shown in \cref{fig:CXpush}(c). For this reason, each Steane block on the Steane layer will have its physical legs arranged in order $4120536$. A similar rotational symmetry of the hypercube defining the QRM code (\cref{fig:Steane-QRM-Stabs}(b)) can be exploited to ensure similar behavior there.

For $\overline{CX}$s straddling pairs of qubits in $\mathcal{Q}$ that are not adjacent, a similar argument applies where we see in \cref{fig:CXpush}(b) that, in the worst case, one physical error spreads to four distinct blocks of Steane codes, which can be corrected in a layer-by-layer error correction scheme. No special care needs to be taken as to how the individual $CX$ gates are implemented. 

\begin{figure}
\centering
\includegraphics[width=0.8\linewidth]{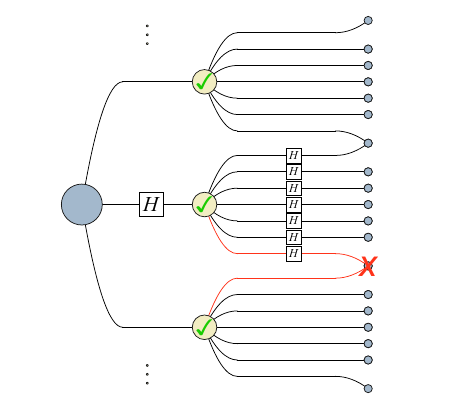}
\caption{Fault tolerance of $\bar{H}$ under operator pushing. A transversal $\bar{H}=H^{\otimes 7}$ in the Steane code (bright yellow disks) is acting on seven QRM codes (dark blue disks) in the new layer. An error on a 2-input QRM code block can propagate to two Steane blocks, an error in each of which is correctable.
}
\label{fig:TandHpushing}
\end{figure}

For targeted $\bar{H}$ on $\mathcal{Q}$, we first implement it transversally on the Steane layer. Then on the QRM outermost layer, the $H$ gates have to be implemented with a Clifford circuit on the $k=1$ QRM blocks. For the $k=2$ QRM block, 
a different circuit is needed to implement $I\otimes H$, but it will be a Clifford circuit, acting only on the present QRM block. For any single qubit error on the $k=1$ QRM blocks, they will contribute to at most one logical error, which is then caught at the Steane layer with error correction. For any single qubit error on the $k=2$ QRM block, the error can propagate into two logical errors because the code can only detect any single qubit error. However, since these two logical errors are propagated into distinct Steane blocks, they are again correctable, as we show in \cref{fig:TandHpushing}.

A similar story holds for the targeted $\bar{T}$ gate on $\mathcal{Q}$, laid out in \cref{fig:pushing.TH.TN}(a) with $T^\dagger$ instead of $T$. First perform the non-transversal circuit for $\bar{T}$ as in \cref{fig:op_pushing}(a), which only acts on three physical qubits in the middle of the Steane block.
This choice of circuit is non-unique, and can be equivalently applied to any three physical qubits $i_1, i_2, i_3$ on which we can find a logical representation $\bar{Z}=Z_{i_1} Z_{i_2} Z_{i_3}$.
Because the circuit consists only of $CX$ and $T$, these can be implemented transversally on the following QRM layer. Any single qubit error on a $k=1$ QRM block on the second layer will always be caught by error correction on the QRM layer. An error on the $k=2$ QRM blocks cannot be corrected, but will again propagate to two distinct Steane blocks, at which point they will be corrected (\cref{fig:TandHpushing}).

\section{Erasure probabilities of seed codes}\label{section:analysis.thresholds}
As discussed in \cref{section:thresholds_for_erasure}, the erasure thresholds and bounds can be estimated once we are given the logical erasure error probability $F(p)$ as a function of the physical erasure rate $p$. For completeness, we provide a more detailed account of their forms for some of the seed codes used in our work.

\subsection{Seed codes from absolutely maximally entangled states (perfect tensors)}

For the $\llbracket 5,1,3\rrbracket$ code which is dual to an AME state (i.e., \emph{perfect tensor}), given some erasure error probability $p$ on the boundary, it is easy to compute the probability of erasure $p_{\text{erasure}}$ of the bulk 
because of the symmetries of the code
\begin{equation}
    p_{\text{erasure}}^{\llbracket 5.1.3\rrbracket}(p) = 1-\big[ (1-p)^5+5(1-p)^4p+10 p^2(1-p)^3\big]~.
\end{equation}
We notice that $p_{\text{erasure}}(1/2)=1/2$ and that $p'_{\text{erasure}}(1/2)>0$. One can also check that this function is monotonically increasing on the interval of $[0,1]$. Therefore, $p_{\text{erasure}}(p)<p$ only when $p<1/2$.

As an exercise, consider joining such seed codes in a regular TTN code encoding, with only one qubit at the center of the bulk. The erasure error probability will only be suppressed when $p<1/2$ and progressively so as we move inwards. It is apparent that this yields an erasure threshold of $p_{th}=1/2$. Returning to asymptotically zero-rate holographic codes, the code can be constructed by concatenating $\llbracket 5,1,3\rrbracket$ and $\llbracket 4,2,2\rrbracket$ codes in various layers depending on the location.

For the $\llbracket 4,2,2\rrbracket$ code obtained by performing \emph{code shortening} \cite{gottesman1997stabilizer} the $\llbracket 5,1,3\rrbracket$ code, it has stabilizer group $S=\langle YXXY, XZZX\rangle$. This shortened code tolerates any single qubit erasure and to access the entire 2-qubit Pauli algebra, we need at least three physical qubits. Any two physical qubits will only support a sub-algebra of the 2-qubit logical Pauli algebra that has support on both logical qubits; this is apparent since any logical operator of the $\llbracket 5,1,3\rrbracket$ code that has a weight of three can be converted into a non-trivial weight-2 logical operator acting on both logical qubits. The erasure probability $p_{\text{erasure}}$ can be obtained similarly as
\begin{equation}
   p_{\text{erasure}}^{\llbracket 4,2,2\rrbracket}(p)= 1-\big[ (1-p)^4+4p(1-p)^3+3(1-p)^2p^2\big]~,
\end{equation}
where for the last term we divided by two, as we only access an Abelian sub-algebra of dimension 2 (instead of 4). Repeating the same analysis above, we get again that $p_{\text{erasure}}(p)$ is monotonically increasing in $[0,1]$ ($dp_{\text{erasure}}/dp\geq 0)$, $p_{\text{erasure}}(1/2)=1/2$ and $p_{\text{erasure}}<p$ if and only if $p<1/2$. Therefore, concatenating these components would on average lead to the same threshold $p_{th}=1/2$ asymptotically. 

In spite of this fact, it is hard to say how erasure recovery is impacted by this sub-algebra in the $\llbracket 4,2,2\rrbracket$ code when one only has access to two of the physical qubits. Therefore, the most conservative estimate is to bound the erasure probability and consider any logical qubit erased unless its entire Pauli algebra is recoverable, ending up with  
\begin{equation}
p_{\text{erasure}}^{\llbracket 4,2,2\rrbracket}(p)\leq 1-\big[ (1-p)^4+4p(1-p)^3\big]~.
\end{equation}
Taking this worst case upper bound, we see that it is still monotonically increasing but $p_{\text{erasure}}>p$ as soon as $p>(5-\sqrt{13})/6\approx 0.2324$, signifying that erasure errors are suppressed only when $p<0.2324$. Indeed, by concatenating $\llbracket 5,1,3\rrbracket$ and $\llbracket 4,2,2\rrbracket$ codes, the erasure threshold should be lower-bounded by $0.2324$. 

\subsection{Seed codes from the Steane code} \label{section:seed.codes.steane.code}

A similar analysis can be performed on the Steane code, wherein we see that 
\begin{align}
    &p_{\text{erasure}}^{\llbracket 7,1,3\rrbracket}(p)
    = 1-\big((1-p)^7 +7p(1-p)^6  \\ &+21 p^2(1-p)^5+ (35-7)p^3(1-p)^4+7p^4(1-p)^3\big)~.\nonumber
\end{align}
This code is relatively easy to count as it is a self-dual CSS code and $\mathsf{\bar{X}}, \mathsf{\bar{Y}}$, and $\mathsf{\bar{Z}}$ all have the same support. Just as before, we obtain a monotonically increasing function with $p_{\text{erasure}}(1/2)=1/2$. 

For the shortened $\llbracket 6,2,2\rrbracket$ code, we again run into a sub-algebra ambiguity, where terms such as $\bar{X}\bar{X}=XXII...,\bar{Z}\bar{Z}=ZZII...$ appear as code sub-algebras supported on two-qubit subsystems;  this implies that they are robust against four qubit erasures. However, the full 2-qubit logical sub-algebra is not recoverable from two physical qubits. A careful counting that includes the sub-algebras as weighted recoverable qubits would yield a crossover point of $1/2$ again by symmetry. 

To get the upper bound on $p_{\text{erasure}}$, we can count the probability that entire logical qubits can be recovered. This also depends on the way the Steane codes are connected together. For now, take the second bulk qubit to be dual to the physical qubit that is described in the center of the triangle, when treating Steane as a two-dimensional color code, 
\begin{align}
    &p_{\text{erasure}}^{\llbracket 6,2,2\rrbracket}(p)\leq 1-\big[(1-p)^6+6p(1-p)^5  \\ &+p^2(1-p)^4(\frac{1}{2}(4)(3)+(4)(3)/2)+p^3(1-p)^3(8/2)\big]~,
    \nonumber
\end{align}
where the first two terms in the square bracket are straightforwardly calculated; the third term counts the number of ways that logical qubit 1 can be recovered. The fourth term counts the number ways by which logical qubit 2 can be recovered if two qubits are erased. Again, to produce a conservative estimate, if the any single logical qubit is not fully recovered (i.e., only an Abelian sub-algebra is erased), then the entire logical qubit is counted as erased. The factor of $\frac{1}{2}$ is to account for the fact that we only recover one logical qubit. Finally, for erasing three qubits, there are eight ways by which the remaining qubits allows us to recover exactly one logical qubit. When erasing more than three qubits, we can only recover Abelian sub-algebras, and so they are not included in the upper bound estimation. , whereas the max rate lower bound is zero.

\section{Details on gate-propagation rules} \label{section:gate.prop.rules}

Here, we show explicitly how the holographic QRM/Steane code allows for gates/operators to be ``pushed'' from one layer of the tensor network to the next. This is necessary not only to show that the entire tensor network encoding map is indeed isometric (which follows from the operator pushing of arbitrary single-qubit Pauli operators), but also that the scheme can be extended to arbitrarily many layers. 

\begin{figure*}
    \centering
    \includegraphics[width=1\linewidth]{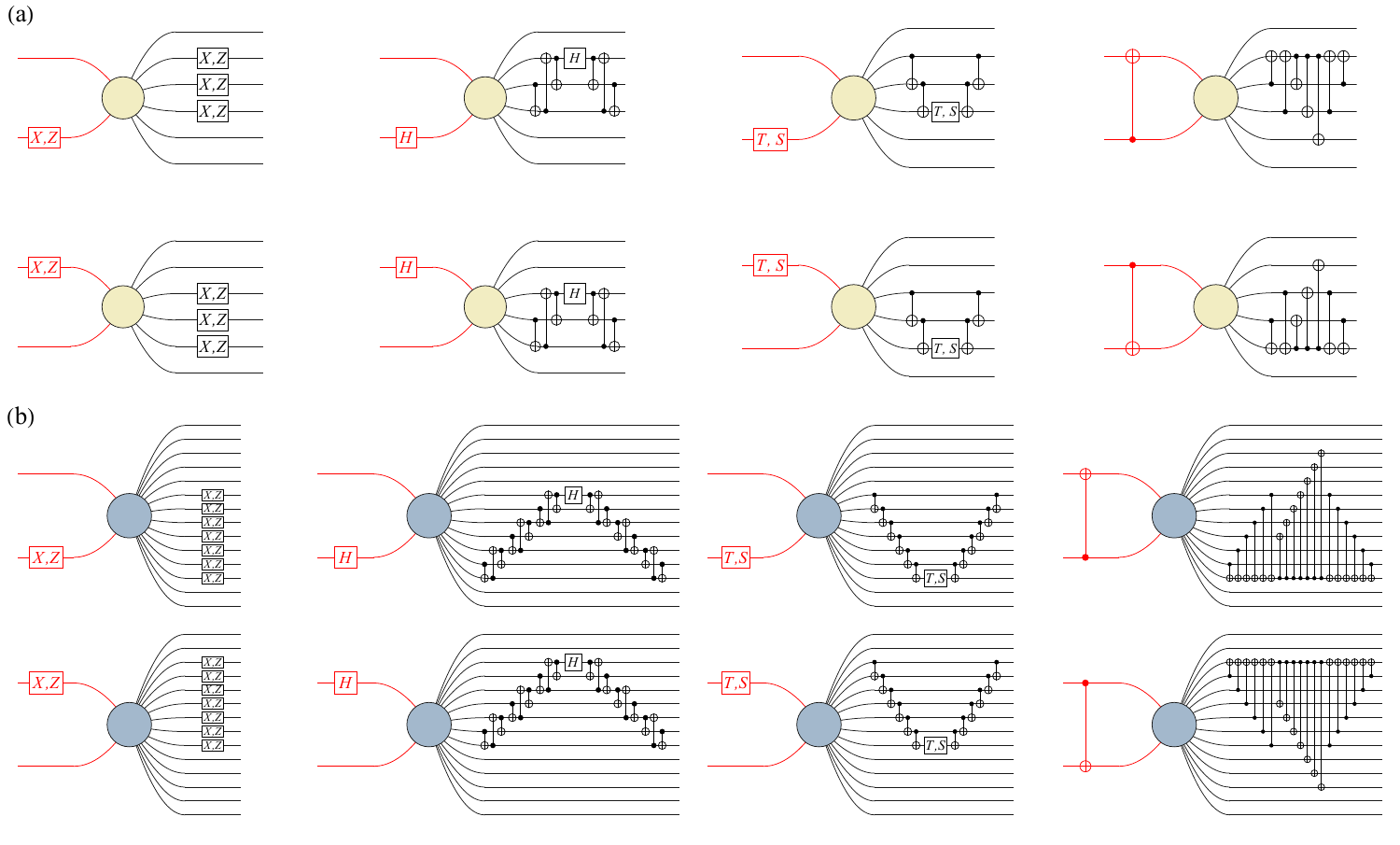}
    \caption{(a) Operator pushing from the logical (red) to physical legs (black) of a modified Steane code with two logical input legs, for the Pauli operators $X$ or $Y$, the phase gate $S$ or Hadamard $H$, the $T$ gate and CX/CNOT 
    gate.
    (b) The same gates of a modified quantum Reed-Muller (QRM) code with two logical input legs. Note that some circuits may have equivalent implementations with fewer gates.
    }
    \label{fig:op_pushing}
\end{figure*}

In \cref{fig:op_pushing}, we consider the case where a Steane or QRM code is used with two logical input legs, which leads to closed loops in the graph geometry, causing deviations from standard tree-style concatenated codes (as shown explicitly in \cref{fig:pushing.TH.TN}). This is the most general case, as it includes operators being pushed from the first input leg, corresponding to the standard Steane/QRM encoding of a single logical qubit, from the second input leg in the two-input configuration, or both. However, in this generality all but the representations of logical Pauli operators will be non-transversal.

The ordering of the physical qubits following the stabilizer definition \eqref{EQ_S_STEANE} and \eqref{EQ_S_QRM} is slightly non-standard to avoid non-transversal gates from acting on the first or last physical qubit, which can cause breakdowns of fault-tolerance when pushing such gates to subsequent layers.

We begin with the pushing rules for logical Pauli operators. For single-input configuration, we have previously seen that $\bar{X}=X^{\otimes n}$ and  $\bar{Z}=Z^{\otimes n}$ are valid representations, with $n=7$ for the Steane code and $n=15$ for the QRM code, respectively.
Using the stabilizers to find a representation that acts trivially on the logical and first physical qubit, we then find the following representation of logical operators for the two-input case,
\begin{align}
    \bar{X}_1 &= XIXIXI \sim IXXXII \ , \\
    \bar{X}_2 &= IXIXIX \sim IIXXXI \ ,
\end{align}
for the Steane code and 
\begin{align}
    \bar{X}_1 &= XIXIXIXIXIXIXI \sim IIIIIXXXXXXXII \ , \\
    \bar{X}_2 &= IXIXIXIXIXIXIX \sim IIXXXXXXXIIIII \ ,
\end{align}
for the QRM code. The subscript denotes which of the two logical inputs the operator acts upon, and the corresponding $\bar{Z}_k$ operators are found by applying $X{\mapsto}Z$ in each $\bar{X}_k$ (note that due to its non-self-dual nature, the QRM code has more representations of $\bar{Z}_k$ than $\bar{X}_k$, which we will not need here). 
Given $\bar{X}_k$ than $\bar{Z}_k$ for each code, we can now derive representations of the $H$, $T$, $S$ and and $CX$ gates. All of these are visualized in \cref{fig:op_pushing}.

The Hadamard gate can be decomposed as $H=(X+Z)/\sqrt{2}$. As we found above, $\bar{X}_k$ and $\bar{X}_Z$ can each be represented on the same $m$ physical qubits, where $m=3$ for the Steane and $m=7$ for the QRM code. It follows that $\bar{H}$ can be represented as $(X^{\otimes m} + Z^{\otimes m})/\sqrt{2}$. To produce such an operator using elementary gates, we note that a single Hadamard can be ``spread'' to two other qubits using by bracketing it with three $CX$ gates
\begin{align}
    &CX_{2,3}\, CX_{3,2}\, CX_{1,2}\, H_1\, CX_{1,2}\, CX_{3,2}\, CX_{2,3} \nonumber\\
    &= \frac{1}{\sqrt{2}} CX_{2,3}\, CX_{3,2}\, (X_1 X_2 + Z_1) \, CX_{3,2}\, CX_{2,3} \nonumber\\
    &= \frac{1}{\sqrt{2}} CX_{2,3}\, (X_1 X_2 + Z_1 Z_3) \, CX_{2,3} \nonumber\\
    &= \frac{1}{\sqrt{2}} (X_1 X_2 X_3 + Z_1 Z_2 Z_3) \ .
\end{align}
Here, we have omitted tensor products and denoted $CX_{j,k}$ as the $CX$ acting on the $k$th qubit controlled by the $j$th. To produce $\bar{H}_k$ for the QRM code, we merely repeat this bracketing twice to reach $m=7$ qubits. Note that this is not the gate-optimal representation of $\bar{H}$: While we require 18 CX gates for the QRM code, Ref.\ \cite{hetero_tree2} includes a more compact circuit in terms of 14. Other representations of logical operators can be derived using the same approach by starting from a different representation of $\bar{X}_k$ than $\bar{Z}_k$.

The $T$ and $S$ gate can each be decomposed into 
\begin{align}
    T &= \frac{1+e^{\ii \pi/4}}{2}\, I + \frac{1-e^{\ii \pi/4}}{2}\, Z \ ,\\
    S &= \frac{1+\ii}{2}\, I + \frac{1-\ii}{2}\, Z \ .
\end{align}
To find a representation of $\bar{T}_k$ and $\bar{S}_k$, we thus have to find a set of gates that produces the above decomposition with $Z \mapsto Z^{\otimes m}$. This can again be achieved using $CX$ gates, e.g., for the Steane code via
\begin{align}
    &CX_{1,2}\, CX_{2,3}\, S_3\, CX_{2,3}\, CX_{1,2}\, \\
    \nonumber
    &= CX_{1,2}\, (\frac{1+\ii}{2}\, I + \frac{1-\ii}{2}\, Z_2 Z_3)\, CX_{1,2} \\
    \nonumber
    &= (\frac{1+\ii}{2}\, I + \frac{1-\ii}{2}\, Z_1 Z_2 Z_3) \ ,
\end{align}
for the $\bar{S}_k$ gate and equivalently for $\bar{T}_k$.

Finally, we consider pushing a $CX$ gate acting on \emph{both} logical inputs. Intuitively, we now need to combine $m$ physical qubits into one effective control qubit that determines an $X^{\otimes m}$ gate on $m$ different physical qubits. Formally, we can expand
\begin{equation}
    CX_{1,2}  = \frac{1}{2} ( I + Z_1 - Z_1 X_2 + X_2 ) \ .
\end{equation}
For simplicity, let us assume 
\begin{equation}
\bar{X}_1 = X_1 X_2 X_3
\end{equation} 
and  
\begin{equation}
\bar{X}_2 = X_2 X_3 X_4
\end{equation} 
and equivalently for $X \mapsto Z$. We then find
\begin{widetext}
\begin{align}
     CX_{2,1}\, CX_{3,1}\, CX_{1,2}\, CX_{1,3}\, CX_{1,4}\, CX_{3,1}\,CX_{2,1}
    &= \frac{1}{2} CX_{2,1}\, CX_{3,1}\, ( I + Z_1 - Z_1 X_2 + X_2 )\, CX_{1,3}\, CX_{1,4}\, CX_{3,1}\,CX_{2,1} \nonumber\\
    &= \frac{1}{2} CX_{2,1}\, CX_{3,1}\, ( I + Z_1 - Z_1 X_2 X_3 X_4 + X_2 X_3 X_4 )\, CX_{3,1}\,CX_{2,1} \nonumber\\
    &= \frac{1}{2} ( I + Z_1 Z_2 Z_3 - Z_1 Z_2 Z_3 X_2 X_3 X_4 + X_2 X_3 X_4 ) \ ,
\end{align}    
\end{widetext}
which ends up in the desired form for $\overline{CX}_{1,2}$. To construct $\overline{CX}_{1,2}$, a similar calculation yields 
\begin{equation}
    CX_{3,4}\, CX_{2,4}\, CX_{4,3}\, CX_{4,2}\, CX_{4,1}\, CX_{2,4}\,CX_{3,4} \ .
\end{equation}
The actual indices for the Steane and QRM codes (for the latter, 18 instead of 7 gates are required) are shown in the last column of \cref{fig:op_pushing}. Again, these gates are decomposed by hand in the simplest manner. Thus optimized implementations with fewer gates are expected to exist.

\end{document}